\documentclass[final]{agujournal2019}
\usepackage{url} 
\usepackage{amsmath}
\usepackage{amssymb}
\usepackage{lineno}
\usepackage{xcolor}
\usepackage{natbib} 
\usepackage{float}
\usepackage{hyperref}
\usepackage{subcaption}
\usepackage{siunitx,  
            booktabs, 
            caption}  
\usepackage{graphicx}
\usepackage{grffile}
\usepackage{makecell}
\usepackage[finalnew]{trackchanges}
\usepackage{nicefrac}
\usepackage{soul}
\usepackage[graphicx]{realboxes}
\journalname{Space Weather}

\begin{document}

%
%


\title{Thermospheric Density, Composition, and Temperature from GOES-R/SUVI Solar Occultations}

%
%




\authors{R. H. A. Sewell\affil{1}, E. M. B. Thiemann\affil{1}, J. Lafyatis\affil{1}, K. Hallock\affil{2}, C. Bethge\affil{2}, M. Pilinski\affil{1}, E. K. Sutton\affil{3}, C. L. Peck\affil{1}, D. B. Seaton\affil{4}}


\affiliation{1}{Laboratory for Atmospheric and Space Physics (LASP), University of Colorado at Boulder, Boulder, CO, USA}
\affiliation{2}{Cooperative Institute for Reasearch in Environmental Sciences (CIRES), University of Colorado Boulder, CO, USA}
\affiliation{3}{Space Weather Technology Research and Education Center (SWx TREC), University of Colorado at Boulder, Boulder, CO, USA}
\affiliation{4}{Southwest Research Institute (SWRI), Boulder CO, USA}




\correspondingauthor{R. H. A. Sewell}{robert.sewell@lasp.colorado.edu}




\begin{keypoints}
\item Measurements of thermospheric O and N$_2$ number densities and neutral temperature are now available from SUVI solar occultation observations.
\item \change{Density comparisons to empirical and assimilative models agree within 15\% at dusk, with larger discrepancies (\mbox{$>$30\%}) at dawn and solar minimum.}{Modeled mass and LOS density comparisons agree within 15\% at dusk but show larger differences (\mbox{$>$20\%}) primarily at dawn and solar minimum.} 
\item \change{These observations began in 2018 and are expected to continue through the mid-18
2030s}{Measurements span +80° to -80° latitude at 6 \& 18 hrs local time during 45 day eclipse seasons around each equinox, starting from Sept 2018.} 
\end{keypoints}

%
%

%
%

\begin{abstract}
\change{A new dataset of atomic oxygen (O) and molecular nitrogen (N$_2$) number density profiles, along with thermospheric temperature profiles between 180 and 500 km, has been developed. These profiles are derived from solar occultation measurements made by the Solar Ultraviolet Imager (SUVI) on the GOES-R satellites, using the 17.1, 19.5, and 30.4 nm channels. Discussed is the novel approach and methods for using EUV solar occultation images for measuring the thermospheric state. Measurement uncertainties are presented as a function of tangent altitude. At 250 km, number density random uncertainties are found to be 8\% and 17\% for O and N$_2$, respectively, and the random uncertainty for neutral temperature at 250 km was found to be 3\%. The impact of effective cross section uncertainty on retrieval bias was assessed, revealing that, as expected, the largest effects occur where O and N$_2$ are minor absorbers. In contrast, total mass density and O/N$_2$ ratios exhibit substantially lower sensitivity, with biases that remain small or nearly constant with altitude. Total mass density comparisons with the NRLMSIS-2.0 model show good agreement at the dusk terminator, with an average difference of -2\%, but larger discrepancies at dawn, with an average difference of -26\%. These discrepancies are more prominent during quiet solar conditions, suggesting an overestimation of densities by MSIS during these conditions. Density comparisons with the IDEA-GRACE-FO and Dragster assimilative models show dawn/dusk percent differences of -24\%/-2\% and +2\%/+13\%, respectively. The dataset is available through the NOAA NCEI GOES-R L2 data pipeline for eclipse seasons from September 2018 onward and is expected to continue through 2035. As this measurement relies only on real-time NOAA space weather SUVI images, these density and temperature profiles could be produced in real-time, supporting critical space weather monitoring and prediction, and filling in a current measurement gap of thermospheric temperature and density.}{A new dataset of atomic oxygen (O) and molecular nitrogen (N$_2$) number density profiles, along with thermospheric temperature profiles between 180 and 500 km, has been developed. These profiles are derived from solar occultation measurements made by the Solar Ultraviolet Imager (SUVI) on the GOES-R satellites, using the 17.1, 19.5, and 30.4 nm channels. Discussed is the novel approach and methods for using EUV solar occultation images for measuring the thermospheric state. Measurement uncertainties are presented as a function of tangent altitude. At 250 km, number density random uncertainties are found to be 8\% and 17\% for O and N$_2$, respectively, and the random uncertainty for neutral temperature is 3\%. The impact of effective cross section uncertainty on retrieval bias was assessed, revealing that the largest effects occur where O and N$_2$ are, respectively, the minor absorber. In contrast, total mass density and O/N$_2$ ratios show substantially lower sensitivity, with biases that remain small or nearly constant with altitude. Total mass density comparisons with the NRLMSIS 2.0 model show good agreement at the dusk terminator (average difference -2\%), but larger discrepancies at dawn (-26\%), particularly during low solar activity. Density comparisons with the IDEA-GRACE-FO and Dragster models show dawn/dusk differences of -24\%/-2\% and +2\%/+13\%, respectively. As this measurement relies only on real-time NOAA space weather SUVI images, these density and temperature profiles could be produced in real-time, supporting critical space weather monitoring and prediction, and helping fill a long-standing observational gap in thermospheric temperature and density.} 
\end{abstract}

\section*{Plain Language Summary}
Extreme ultraviolet (EUV) light from the Sun is strongly absorbed by the upper atmospheric layer of Earth, the thermosphere. When measuring the attenuation of light over different EUV wavelengths, the density, composition and temperature of the thermosphere can be inferred. Presented here is the novel method to make measurements of atomic oxygen (O), molecular nitrogen (N$_2$) and temperature in the thermosphere using solar EUV images from the Solar Ultraviolet Imager (SUVI), on the Geostationary Operational Environmental Satellite-R Series (GOES-R) satellites, during times when the imager passes into and out of Earth's shadow, known as solar occultations. These new measurements show reasonable agreement with data from existing empirical and data assimilation models, but some discrepancies are observed, particularly at dusk and during low solar activity. Because there are so few measurements of this type currently being made, these measurements are useful for space weather nowcasting modeling validation and assimilation.

%
%

%


%
%
%
%

\section{Introduction}

The Earth's thermosphere, stretching from the mesopause to the exobase ($\sim$ 85 km - 500 km), is a critical region of the upper atmosphere that plays a pivotal role in space weather and satellite operations. These altitudes include both the ionosphere and the Low Earth Orbit (LEO) region, and its state is influenced by a variety of photochemical and dynamical processes. These processes \change{effect}{affect} the thermosphere's temperature, density and composition which directly impact satellite drag and -- through ionosphere-thermosphere (IT) coupling -- the propagation of electromagnetic signals through the ionosphere. The thermospheric state (density, temperature and composition) variations are driven largely by variability in the solar extreme ultraviolet (EUV) and soft X-ray (SXR) irradiance \mbox{\citep{Solomon2005}}, which \change{is}{are} the primary heating \change{source}{sources} of the thermosphere. However, the thermosphere is also influenced from above by magnetic and electrical energy and energetic particles from the magnetosphere, as well as below through waves and tides originating in the lower atmosphere, all resulting in complex spatial and temporal structure \mbox{\citep{SOLOMON2015402}}. Despite the importance of studying the neutrals in this region for understanding atmospheric dynamics and their impact on space weather, few \add{operational} measurements exist\add{,} \remove{-- with none being made for real-time operational purposes -- }leading to this region being dubbed the ``Thermospheric Gap" \mbox{\citep{Oberheide2011, Forbes2022, Jones2022}}.

Historically, the study of the thermosphere has relied on a variety of techniques and instrumentation. Early in situ measurements of neutral species were made by the Atmospheric Explorer E (AE-E) and Dynamics Explorer (DE) missions using mass spectrometry  \citep{Carignan1981, Nier1973}. These mission had highly elliptical orbits which enabled in-situ measurements of neutral densities of multiple atmospheric species and neutral temperature over a range of altitudes in the thermosphere. Other in situ measurements, such as the Challenging Minisatellite Payload \citep[CHAMP; ][]{Bruinsma2003}, the Gravity Recovery and Climate Experiment \citep[GRACE; ][]{Sutton2007}, the Gravity Field and Steady-State Ocean Circulation Explorer \citep[GOCE; ][]{Bruinsma2014}, and Swarm \citep{Siemes2016} have measured total mass density at the respective satellite's orbit altitude using precision accelerometer data. The Thermosphere, Ionosphere, Mesosphere Energetics and Dynamics (TIMED) mission's Global Ultraviolet Imager (GUVI) provided altitude-resolved measurements of \add{O, N$_2$, O$_2$, and thermospheric temperature from} atmospheric Far Ultraviolet (FUV) airglow \remove{through }limb scans until 2008\remove{,} when a mechanism failure ended this capability \citep{Meier2015}. Since then, TIMED/GUVI provides \add{these} \remove{column-integrated airglow }measurements without altitude resolution. More recently, the Ionospheric Connection Explorer (ICON) mission provided daytime column-integrated O/N$_2$ density ratios \citep{Meier2021} and limb O and N$_2$ profiles \citep{Stephan2018} from FUV dayglow measurements, until contact was lost with the satellite in November of 2022. In addition, the Global-scale Observations of the Limb and Disk (GOLD) mission uses FUV remote sensing techniques to measure altitude-resolved O$_2$ density using stellar occultations \citep{Lumpe2020}, column-integrated O/N$_2$ and effective disk temperature near the airglow peak ($\sim$160 km) from airglow measurements \citep{Eastes2017, Correira2021}, and neutral temperature (at 250 km) from airglow measurements \citep{Evans2020}.

\change{Because EUV radiation is strongly absorbed in the thermosphere,}{In addition to these missions and measurement methods,} full-disk EUV irradiance solar occultations have also been used previously for studying density and composition in the upper atmosphere. In recent decades, no dedicated instruments have been launched solely for solar occultation observations of the thermosphere. Instead, most \change{such}{recent} measurements have been \change{carried out as ``bonus" science}{opportunistic in nature, using data from}\remove{ by} instruments originally \add{designed} for solar research. These measurements include O$_2$ density profiles from the Ultraviolet Spectrometer and Polarimeter on the Solar Maximum Mission \citep[SMM/UVSP;][]{Aikin1993} and from Solar Ultraviolet Spectral Irradiance Monitor aboard the Upper Atmospheric Research Satellite \citep[SUSIM/UARS; ][]{Lumpe2007}, as well as O and N$_2$ densities and temperature from the Large Yield Radiometer on the Project for Onboard Autonomy 2 \citep[PROBA2/LYRA; ][]{Thiemann2017, Thiemann2021}. Despite the scarcity of such measurements, solar occultations are a useful complement to other measurement techniques in addressing the  ``Thermospheric Gap". In comparison to in-situ accelerometer measurements\remove{ or column integrated airglow measurements}, occultation measurements can provide altitude resolved profiles of the thermospheric state.\add{ Occultation measurements, in comparison to FUV airglow measurements, do not rely on precise knowledge of the primary radiance measurement and instead rely on the ratio of occulted to non-occulted irradiance, allowing for measurements to be self-calibrating. Furthermore, solar occultations, in particular, provide measurements at the solar terminator, a region where FUV airglow measurements are difficult to interpret due to slant path anisotropy of the solar illumination.} 

In this paper, new measurements of the thermospheric density\add{,} temperature and composition from 180 to 500 km are presented. These are derived from solar occultation observations by the Solar Ultraviolet Imager (SUVI) \citep{Vasudevan_2019, Darnel2022} on the Geostationary Operational Environmental Satellites (GOES)-R Series satellites \citep{Goodman2019}. These measurements yield \remove{1.5 km}vertically resolved density profiles of atomic oxygen (O) and molecular nitrogen (N$_2$) and neutral temperature. Also described are the novel methods for using EUV solar occultation images for thermospheric measurements, in contrast to previous full-disk irradiance measurement techniques \citep{ROBLE19721727,Thiemann2017}. The SUVI Level 1b (L1b) solar images are a near real-time data product with $\sim$1 minute latency, however this occultation dataset is being produced as part of the nominal SUVI level 2 (L2) science data pipeline, which has a typical latency of 2 days. The use of L2 data instead of L1b data is solely due to funding constraints which have prevented the occultation data processing code from being rewritten in a manner that is compliant with the more restrictive real-time data processing environment. These new observations provide critical measurements of the Thermospheric Gap, allowing for a better understanding of the complex coupling in the thermosphere, as well as allowing for better model prediction and validation for use in space weather operations. As the generation of this dataset only relies on the real-time space weather SUVI solar images, these density and temperature profiles have the potential to be produced \change{in real-time for space weather applications.}{as an additional SUVI space weather product, with a latency on the order of $\sim$10 minutes after an occultation begins. The occultation duration themselves are 7 minutes and the data processing completes in approximately 3 minutes.}

\section{Methods and Data}
Solar occultation measurements are made using the SUVI instruments on board the National Oceanic and Atmospheric Administration (NOAA) GOES-R series satellites (GOES-16, GOES-17, GOES-18, and GOES-19). SUVI images the Sun over six distinct EUV spectral passband channels (9.4, 13.1, 17.1, 19.5, 28.4, and 30.4 nm) to distinguish solar features of interest, utilizing a 4-minute imaging sequence of long (1 s) and short (5 ms) exposure times over all six bands (see \citet{Vasudevan_2019}, Figure 3 for the exact channel measurement sequence). The methods presented here for resolving thermospheric temperature, density, and composition utilize the long exposure 17.1, 19.5 and 30.4 nm channel images as they provide adequate signal-to-noise during all solar activity levels\remove{, for the data processing methods described in Sections {\ref{subsec:image_reg}} \& {\ref{subsec:forward_model}}}. The average occultation duration, between when the EUV signal begins to attenuate to when it is fully attenuated, is roughly 3 minutes\change{; this means}{. As such,} the sampling of these long exposure 17.1, 19.5, and 30.4 nm images varies between occultation observations\remove{,} depending on where the occultation falls in the nominal 4-minute imaging sequence\change{,}{.} \change{at least two distinct measurements from different channels}{However, each observation will include distinct measurements from at least two different channels}. The density and temperature product described here can be produced with any combination of these three images observed during a given occultation. The first GOES-R series satellite (GOES-16) became operational in 2017, while the last (presumably, GOES-19) is expected to be operational through 2036\change{, thus}{. Thus} the future SUVI data record will span nearly two solar cycles. At the time of this writing, GOES-16 and GOES-18 are operational\remove{,} with the remaining two GOES-R spacecraft in on-orbit storage.

\begin{figure}[H]
    \centering
    \includegraphics[width=0.75\linewidth]{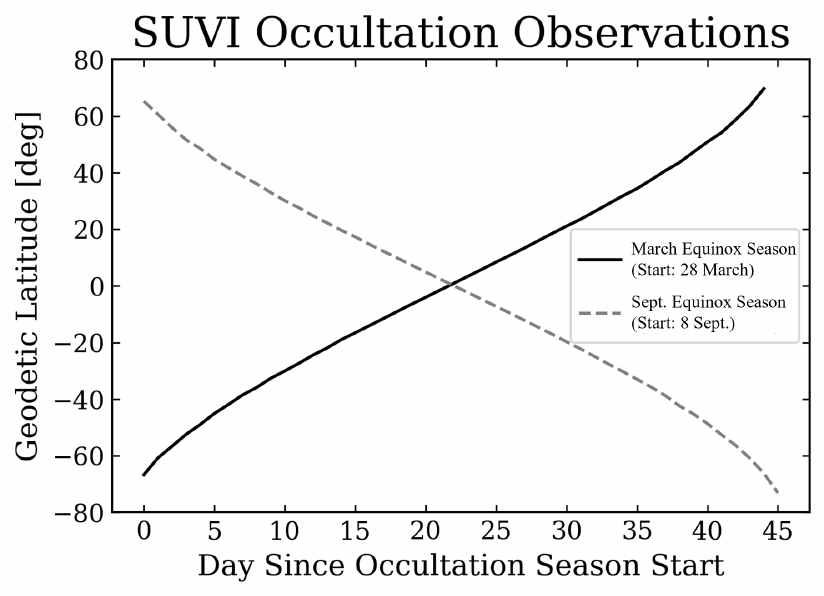} 
    \caption{During a 45-day occultation season, occurring around the March and September equinoxes of each year\remove{(approximate start date given as DD/MM)}, SUVI occultations on a given GOES-R satellite will cover nearly pole-to-pole with two local time measurements made at each latitude, one at the dusk and one at the dawn terminator.}\label{fig:occ_lats}
\end{figure}

\begin{figure}[H]
    \centering
    \includegraphics[width=\linewidth]{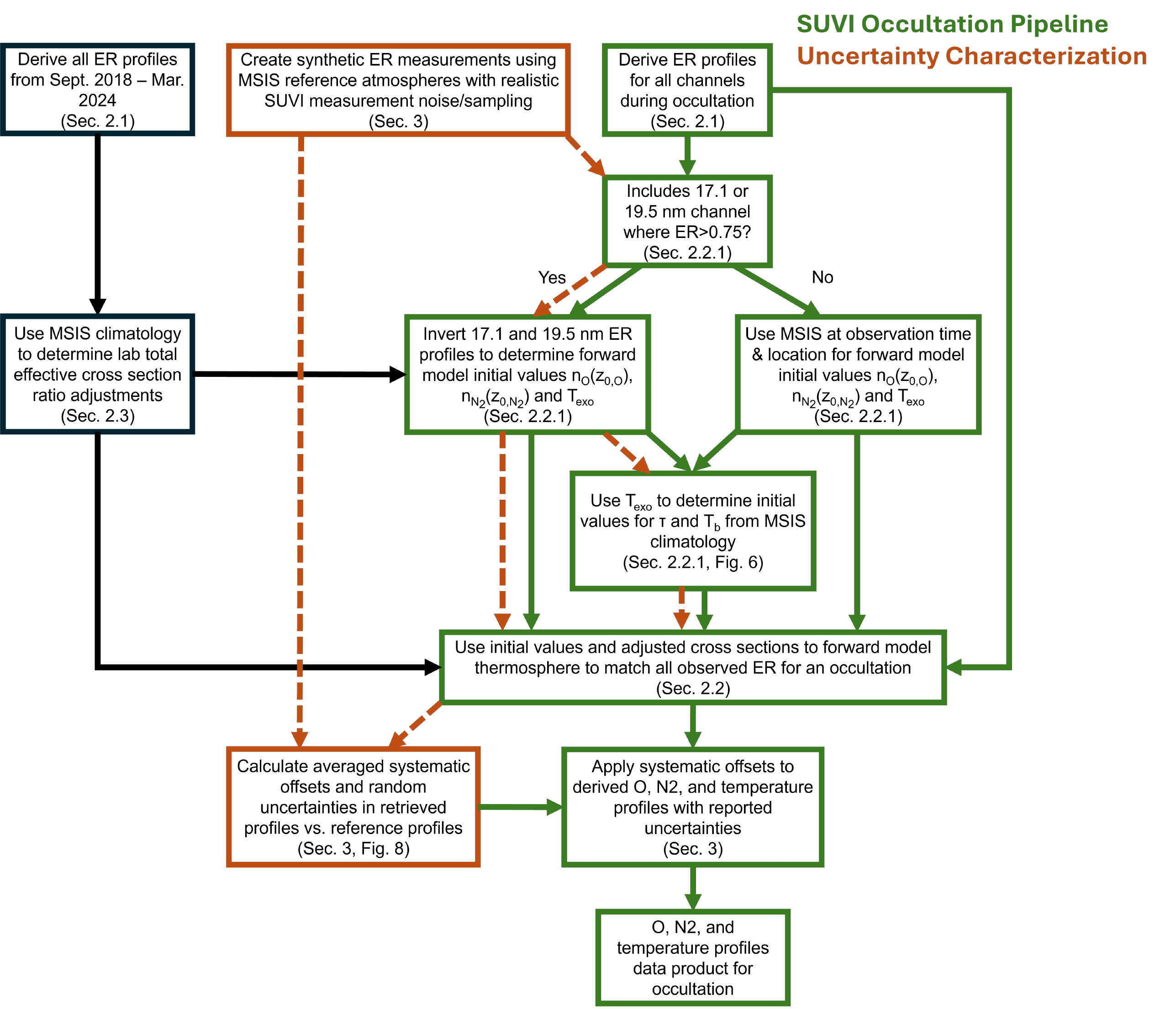} 
    \caption{Flowchart of SUVI solar occultation retrieval algorithm.}\label{fig:flowchart}
\end{figure}

The GOES-R series satellites operate in geostationary orbit near either the GOES-East (75\textdegree W) or the GOES-West (135\textdegree W) slots. \change{These orbits experience eclipse seasons centered around the March and September equinoxes, lasting each for $\sim$45 days. During these eclipse seasons, the observing latitude changes from orbit-to-orbit, scanning from pole-to-pole over the 45-day season, as shown in Figure \mbox{\ref{fig:occ_lats}}. During these periods, twice per day, as the spacecraft enters eclipse at the dusk terminator and exits eclipse at the dawn terminator, SUVI images the Sun through Earth's atmosphere.}{These orbits have $\sim$45-day eclipse seasons around the March and September equinoxes. The observing latitude shifts each orbit, scanning from pole to pole as shown in Figure \mbox{\ref{fig:occ_lats}}. Twice daily the spacecraft enters eclipse near the dusk terminator and exits near the dawn terminator, allowing SUVI to image the Sun through Earth's atmosphere.} Because the GOES-East and GOES-West positions are above different longitudes, consecutive dawn-dawn or dusk-dusk occultation measurements by SUVI  on each satellite will observe roughly at the same latitude but separated by $\sim$4 hours. \add{Due to their geostationary orbit, GOES-East dawn observations occur roughly over 5\textdegree E and dusk observations occur over 150\textdegree W longitude, while GOES-West dawn observations occur over 60\textdegree W and dusk observations occur over 145\textdegree E longitude.}
The solar radiance measured by SUVI during these occultation times is attenuated as a function of the composition and density of the horizontal atmospheric line-of-sight (LOS) column between the spacecraft and the Sun. The fundamental measurement of these observations is given by a form of the Beer-Lambert law as
\begin{equation}\label{eqn:beer_lambert}
    \frac{I_\lambda}{I_{0,\lambda}}=\exp\left(-\sum_{i}N_i\sigma_{i,\lambda}\right)
\end{equation}
where $I_\lambda$ and $I_{0,\lambda}$  are, respectively, the attenuated and non-attenuated solar irradiance measured over a given passband \add{(in units of W/m$^2$)}, $N_i$ is the integrated horizontal column LOS density \add{(in units of cm$^{-2}$)}, hereafter referred to as the LOS density, of the $i$th absorbing species, and $\sigma_{i,\lambda}$ is the total effective absorption cross section of the $i$th species \add{(in units of cm$^{2}$)} weighted by the spectral response of the passband. \change{By co-registering pixels in SUVI pixels in the occulted images with the temporally nearest non-occulted pixels, the attenuation or extinction ratio (ER), $I_\lambda/I_{0,\lambda}$, can be calculated for each channel during an occultation, as discussed in Section {\ref{subsec:image_reg}}}{Using SUVI solar images taken during an occultation and the temporally nearest non-occulted image, the attenuation $I_\lambda/I_{0,\lambda}$, or the extinction ratio (ER), can be derived}. When multiple channel measurements overlap to observe the same atmospheric LOS column, their ERs can be used to distinguish between absorption due to \change{O and N$_2$ densities of the LOS column}{the LOS column densities of the thermospheric major species, O and N$_2$,} \change{if the cross sections $\sigma_{O,\lambda}$ and $\sigma_{N_2,\lambda}$ are unique between the observation passbands}{if cross sections for an individual species are unique between the observation passbands}. Furthermore, by modeling the atmosphere that would produce such LOS densities\remove{, as discussed in Section {\ref{subsec:forward_model}}}, \add{ER} measurements at different altitudes can be related to one another, thus improving uncertainties\change{, as well as}{ and} allowing a temperature profile to be \change{developed}{derived} from the data.\add{ Figure {\ref{fig:flowchart}} outlines the retrieval algorithm used to process SUVI occultation images and derive thermospheric parameters, along with the corresponding sections that describe each step in detail. Section {\ref{subsec:image_reg}} discusses the SUVI image pre-processing required to obtain the extinction ratio (ER) measurements. Section {\ref{subsec:forward_model}} presents the forward modeling approach and the methods used to determine initial values for retrieving thermospheric density and temperature. Section {\ref{subsec:CS_Uncertainty}} describes the determination of absorption cross sections used in the inversion process. Finally, Section {\ref{sec:error_fwd_model}} addresses the uncertainty analysis and the application of standard profile adjustments.}
\subsection{Solar Image Pre-processing \& Extinction Ratio Generation}\label{subsec:image_reg}

\begin{figure}[H]
    \centering
    \includegraphics[width=.75\linewidth]{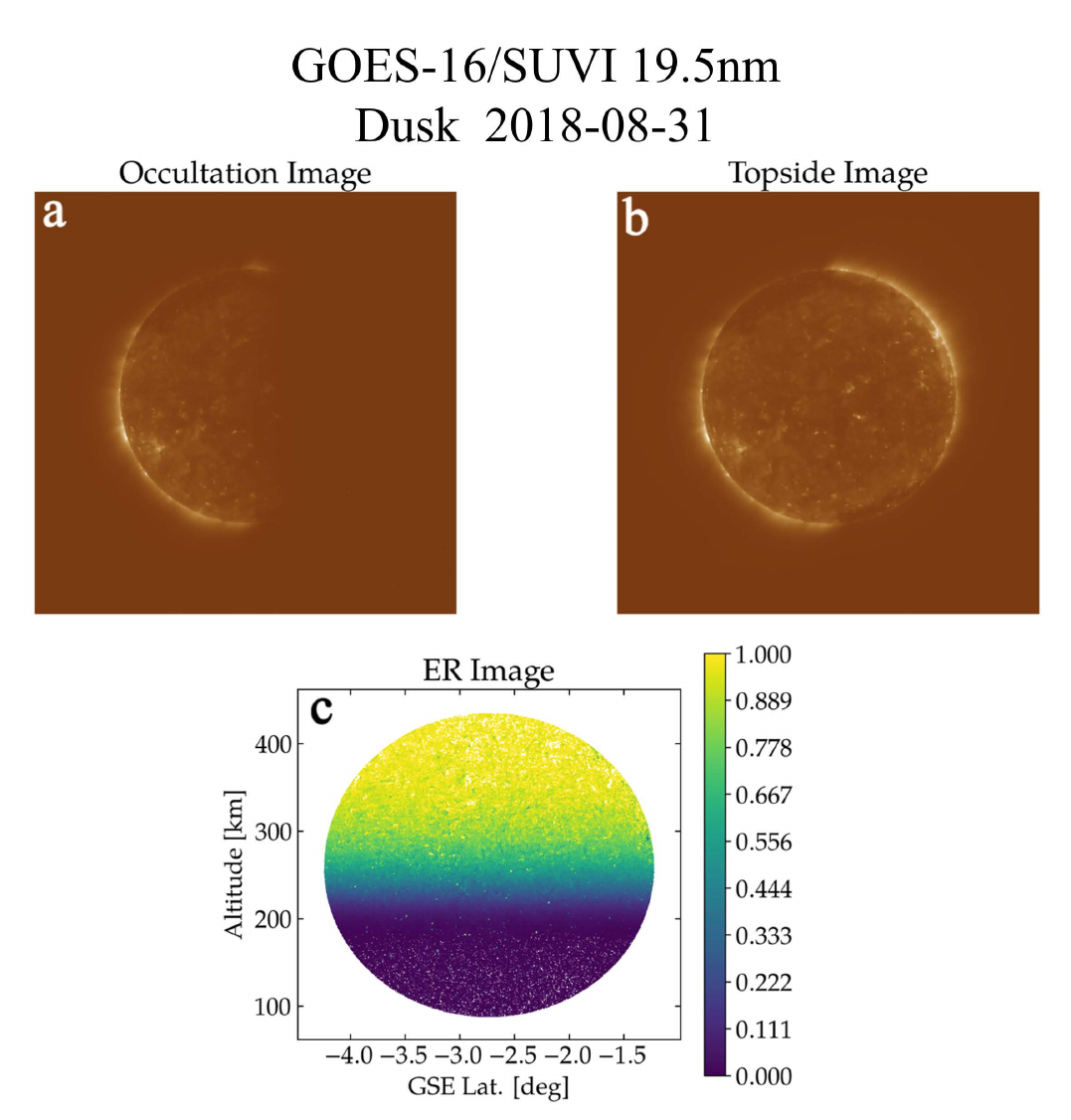} 
    \caption{\change{Example occultation (a) and topside (b) image pair for a SUVI solar occultation measurement with resulting extinction ratio image (c) after phase cross-correlation and terminator plane projection.}{Example raw SUVI L1b occultation (\textbf{a}) and topside (\textbf{b}) images, with resulting extinction ratio image (\textbf{c}) after phase cross-correlation, terminator plane projection, and rotation.}}\label{fig:occ_top_er_images}
\end{figure}

\begin{figure}[H]
    \centering
    \includegraphics[width=.65\linewidth]{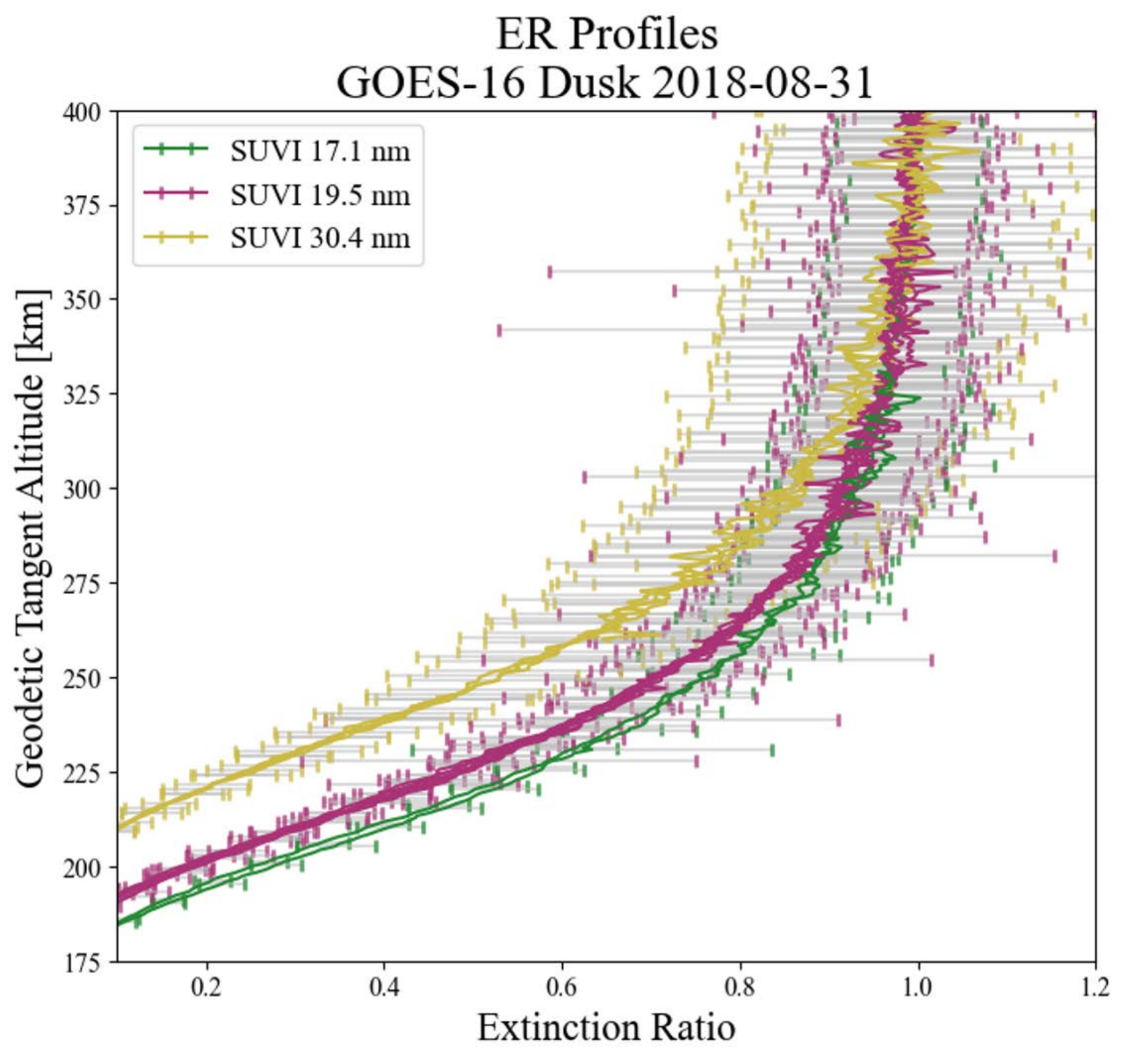} 
    \caption{Example ER profiles for all occultation images taken during the dusk 2018-08-31 SUVI observation\add{, with associated 1-$\sigma$ uncertainties in gray with endcap colors corresponding to their associated channels.}}\label{fig:er_profile}
\end{figure}

\change{At the altitude of geostationary orbit, an image of the solar disk, when projected into the terminator plane at Earth, spans roughly 300 km.}{From geostationary orbit, the projection of the solar disk into the terminator plane at Earth spans roughly 300 km.} The \add{tangent} altitude at which solar EUV light absorption begins to be perceptible during a SUVI solar occultation typically begins around \remove{an altitude of }400-500 km, with signals being totally extinguished \remove{by }around 180 km\change{, over the course of an $\sim$3 minute occultation observation}{. This occultation observation period typically lasts $\sim$3 minutes}. In order to generate the ER for each channel, the LOS tangent ray height\remove{, $z_t$,} of each image's center pixel\add{, $z_t$,} is calculated using the satellite and Sun location in the Geocentric Solar Ecliptic (GSE) reference frame. Occultation images are then identified such that $700\ km>z_t>50\ km$.  Furthermore, the \change{most recent}{temporally nearest} non-occulted image taken on the same channel, henceforth referred to as the topside image, is identified such that $z_t > 1000\ km$. \add{These occultation and topside images are taken within, at most, 7 minutes of one another to minimize potential changes on the solar disk.} The pixels of the occultation image are then co-registered to that of the corresponding channel topside image. During occultations, \change{it was found that the solar pointing pixel information can drift from the true telescope pointing}{the reported Sun-center pixel coordinate in the SUVI L1b data can differ from what appeared to be true Sun-center, due to occlusion of the solar disk}, so this co-registration is done independently through fast Fourier transform phase cross-correlation \add{\mbox{\citep{fft_cross}}}. \add{Using solar disk features that are not occulted, this cross-correlation provides the sub-pixel level vertical and horizontal shift required to align the solar disk in the occultation image to that in the topside image.} The ratio of the co-registered occultation image to the topside image, results in an extinction ratio image (ER image). An example of an identified occultation, topside and ER image for a SUVI \change{17.1}{19.5} nm occultation observation is shown in Figure \ref{fig:occ_top_er_images}.
To register these ER image pixel values to the LOS column of \add{the} atmosphere being measured, the pixel mapping of the solar disk and the satellite location at the time of measurement (both provided in the GOES-R/SUVI L2 data files) are used to project the pixel map into the tangent plane, yielding a three-axis geodetic coordinate of the tangent point for each ER pixel's line-of-sight. This projection is illustrated in Figure \ref{fig:occ_top_er_images}c showing altitude and GSE latitude map of an example ER image. \add{The spatial sampling of each pixel is roughly 0.5 km with providing a spot width resolution of $1.5\pm0.5$ km.}
\change{The ER image pixel values are then binned into 0.5 km bins and averaged according to their tangent ray height, calculated from the pixel's GSE terminator projection location, to provide better signal-to-noise, resulting in a vertical ER profile at an average latitude of the observation.}{All ER image pixel values are then averaged and binned into 0.5 km bins according to their GSE tangent ray altitude, resulting in a vertical ER profile at the average observation latitude.} \add{While the ER images are sensitive to transient events on the Sun at the pixel level -- e.g. flares -- these events are not identified or removed from the images. Instead, the altitude binning of the ER image into the ER profile reduces the impact of small transients on the Sun. Furthermore the standard deviation of the binned ER pixels provide the ER profile uncertainty, which is used to inform the forward model described in the following section.} These ER profiles are generated for all measurements of the 17.1, 19.5 and 30.4 nm channels for a given occultation, as shown by example in Figure \ref{fig:er_profile}, which are then used to derive neutral temperature and O and N$_2$ \remove{LOS} densities.
\subsection{Thermosphere Forward Model}\label{subsec:forward_model}
From the ER profiles derived through the methods above, Equation \ref{eqn:beer_lambert} can be inverted to solve directly for the LOS densities of O and N$_2$ at each measurement altitude where at least two channels have measurements. Alternatively, by modeling the thermospheric state that would produce such ER profiles, measurements at different altitudes and channels can be related to better constrain the density measurements of a given occultation.\add{ The primary trade-off in modeling the thermosphere using this approach is the loss of vertical resolution, as the equations of state represented in Equations {\ref{eqn:bates}} and {\ref{eqn:diffusive_equalibrium}} inherently smooth vertical profiles. Although a direct inversion of SUVI ER profiles (see Figure {\ref{fig:direct_cs}}) or ER images \mbox{\citep{Thiemann2023}} could, in principle, be used to probe vertical structure, this study adopts a forward modeling approach that does not treat structure beyond diffusive separation (e.g. gravity waves). Densities and temperatures are retrieved using the full vertical extent of each occultation, as described in this section.} 

The equation of state used to describe the thermospheric temperature in this algorithm is the \change{Bate's}{Bates} temperature profile \citep{Bates1959}, given as a function of geopotential height:
\begin{equation}\label{eqn:bates}
    T(\zeta)=T_{exo}-(T_{exo}-T_b)\exp\left[-\tau(\zeta-\zeta_b)\right]
\end{equation}
 where $T_{exo}$ is the exospheric temperature \add{(in Kelvin)}, $T_b$ is the temperature \add{(in Kelvin)} at a reference height ($\zeta=\zeta_b$), $\tau$ is the shaping parameter defined as  
\begin{equation}\label{eqn:tau}
    \tau=\frac{\left(\frac{dT}{d\zeta}\right)_{\zeta=\zeta_b}}{(T_{exo}-T_b)}
\end{equation}\label{fig:bates}
and where $\zeta$ is the geopotential height defined as
\begin{equation}\label{eqn:geopentential}
    \zeta=\frac{1}{g_0}\int_{0}^zg(z)dz
\end{equation} 
where $g$ is the gravitational acceleration from Earth \add{(in units m/s$^{2}$)}, $g_0$ is the mean gravitational acceleration at sea level, and z is the tangent point height of interest \add{(in meters)}.

For the number density of a neutral species in the thermosphere \add{(in units of cm$^{-3}$)}, it is common to assume the vertical profile will follow that of diffusive equilibrium, given by  
\begin{equation}\label{eqn:diffusive_equalibrium}
    n_s(z)=n_s(z_{0,s})\frac{T(z_{0,s})}{T(z)}\exp{\left(-\int_{z_{0,s}}^z\frac{dz}{H_s(z)}\right)}
\end{equation}
where $T$ is the \change{Bate's}{Bates} temperature profile from Equation \ref{eqn:bates}, \change{$z_0$}{$z_{0,s}$} is a reference altitude above Earth's surface\add{ for a given species}, and $H_s$ is the species scale height
\begin{equation}\label{eqn:scale_height}
    H_s(z)=\frac{k_BT(z)}{m_sg(z)}
\end{equation}
for $m_s$ the mass of the species \add{(in kilograms)}, $g$ \change{Earth's gravitational coefficient}{the gravitational acceleration from Earth}, and $k_B$ Boltzmann's constant \add{(in units of J/K)} \citep{Schunk_Nagy_2009}. Assuming spherical symmetry, this number density is related to the LOS density of a given species through the Abel transform \citep{Abel1826, ROBLE19721727} as 
\begin{equation}\label{eqn:abel_transform}
    N_s(z)=2\int_z^\infty\frac{n(r)rdr}{\sqrt{r^2-z^2}}
\end{equation}
where r is the radial altitude \add{(in meters)} of each point along the LOS track. From Equation \ref{eqn:beer_lambert}, the ER profile for a given occultation can be derived using the species LOS densities of all absorbing species in the LOS column. 

\begin{figure}[H]
    \centering
    \includegraphics[width=.95\linewidth]{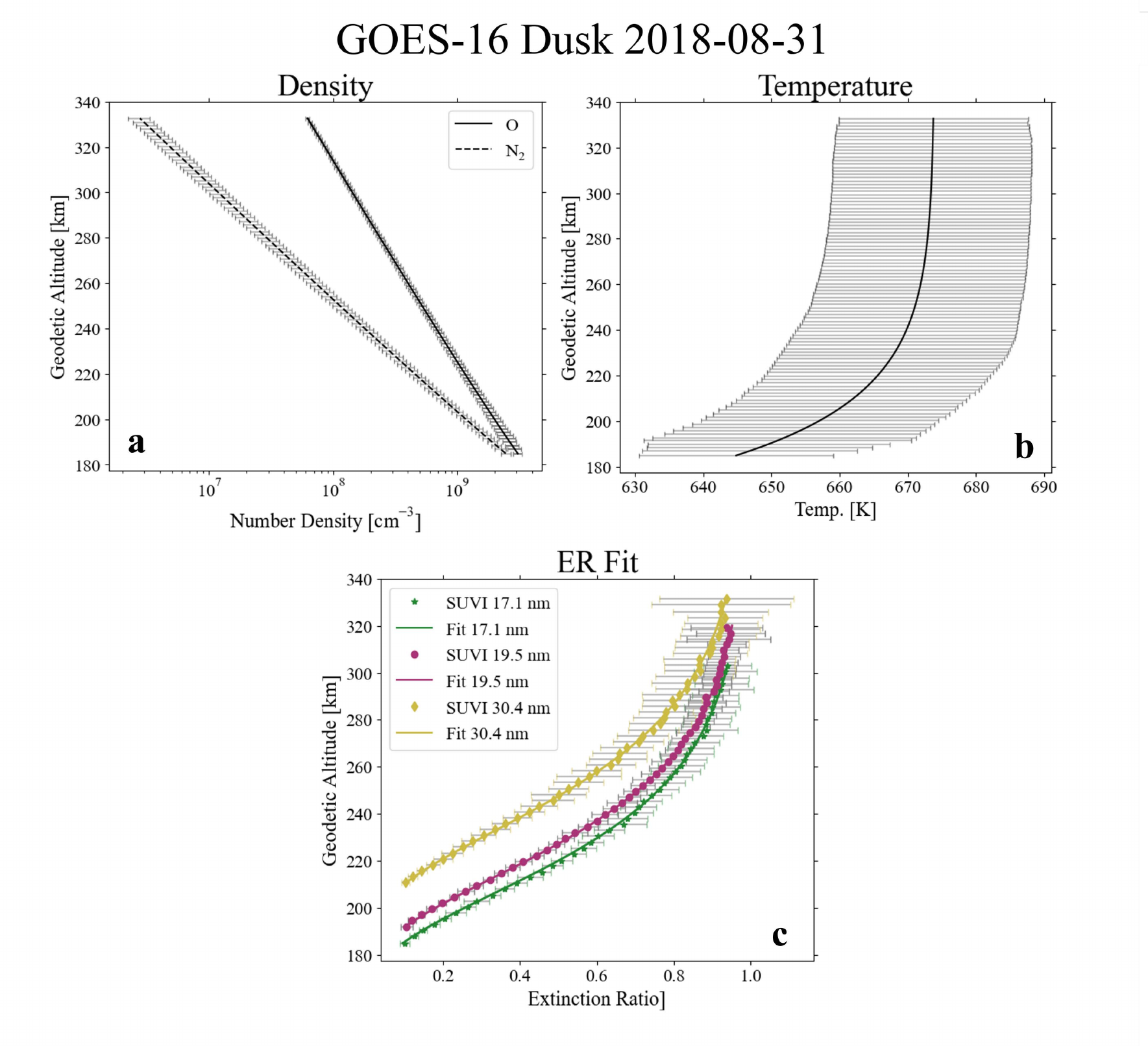} 
    \caption{\textbf{(a)} Example SUVI derived O and N$_2$ number density profile and \textbf{(b)} temperature profile from the dusk 2018-08-31 SUVI solar occultation observation\add{, with random uncertainties marked in gray as derived in Section~{\ref{sec:error_fwd_model}}}. \textbf{(c)} Fitted atmosphere generated ER profiles compared to SUVI measured ER profiles. \add{The SUVI measured average ER uncertainties in \textbf{c} are marked in gray with endcap colors matching the corresponding channel.}}\label{fig:den_and_temp}
\end{figure}

From equations of state, Equations \ref{eqn:bates} \& \ref{eqn:diffusive_equalibrium}, thermospheric ERs can be modeled using Equations \ref{eqn:abel_transform} \& \ref{eqn:beer_lambert} (assuming a thermosphere dominated by atomic oxygen and molecular nitrogen) with the five state variables: $T_{exo}$, $T_b$, $\tau$, \change{$n_o(z_0)$}{$n_o(z_{0,o})$}, and \change{$n_{n_2}(z_0)$}{$n_{n_2}(z_{0,n_2})$}. These variables are forward modeled to match the generated ER profiles of each channel across all measured altitudes for each occultation observation. This is done using a Levenberg-Marquardt (LM) non-linear least squares minimization algorithm implemented in the publicly available LMFIT python package \citep{newville_2015_11813}. Through optimizing these parameters to the ER data, full vertical thermospheric temperature ($T(z)$), number density ($n_o(z)$ \& $n_{n_2}(z)$), and LOS density ($N_o(z)$ \& $N_{n_2}(z)$), profiles are derived. \add{While these vertical profiles are valid over all altitudes where the assumption of diffusive equilibrium holds, the reported profiles only span the altitudes where there was ER profile coverage.} Figure \ref{fig:den_and_temp} shows an example forward model output and SUVI solar occultation data product for the O and N$_2$ number density and neutral temperature from the 2018-08-31 dusk observation. \add{Uncertainties for the temperature profile and the O and N$_2$ number densities are described in Section \mbox{\ref{sec:error_fwd_model}}.}

\subsubsection{Forward Model Parameter Initial Values}\label{sec:seed_values}
For forward modeling the SUVI occultation observations, initial values of the state variable T$_{exo}$, T$_b$, $\tau$, \change{$n_O(z_0)$}{$n_O(z_{0,O})$}, and \change{$n_{N_2}(z_0)$}{$n_{N_2}(z_{0,N_2})$} must be determined. 

\change{The absorption cross sections for O and N$_2$ across the 17.1 nm and 19.5 nm passbands are nearly equal \mbox{\citep{Thiemann2017}}; i.e., $\sigma_{O,17.1}\approx\sigma_{N_2,17.1}$ and $\sigma_{O,19.5}\approx\sigma_{N_2,19.5}$. This fact can be leveraged to derive initial values of T$_{exo}$, $n_O(z_{0,O})$ and $n_{N_2}(z_{0,N_2})$ by assuming these cross sections to be exactly equal, and inverting Equation {\ref{eqn:beer_lambert}} to find the total O(z)+N$_2$(z) LOS density for the available 17.1 and 19.5 nm altitude averaged ER profiles. At high altitudes, where the $ER>0.75$, O is expected to be the major species and the temperature is nearly or completely isothermal. As such, an exponential fit of the O(z)+N$_2$(z) LOS density profile provides an initial estimate of the O-only LOS density at high altitudes. This O LOS density is then inverted, using the  inverse Abel transform of Equation {\ref{eqn:abel_transform}}, to derive an estimate of the initial number density seed value $n_O(z)$. A reference altitude of $z_{0,O}=300$ km is used to then provide the initial value $n_O(z_{0,O})$. Because at high altitudes the temperature is expected to be near isothermal, the slope of a line fitted to the natural logarithm of the previously derived high altitude profile of $n_O(z)$ will be approximately the scale height of atomic oxygen (Equations {\ref{eqn:diffusive_equalibrium}} \& {\ref{eqn:scale_height}}). From this scale height, the initial value of T$_{exo}$ is derived. Furthermore, the residual of the measured O(z)+N$_2$(z) total LOS density and fitted O(z) LOS density, provides an initial estimate of N$_2$(z) LOS density. This altitude resolved N$_2$(z) LOS density estimate is passed through the inverse Abel transform (inverse of Equation {\ref{eqn:abel_transform}}) to derive an estimate of the density for N$_2$, $n_{N_2}(z)$. Using this number density profile at a reference altitude of $z_{0,N_2}=250$ km provides the initial value for $n_{N_2}(z_{0,N_2})$. If no 17.1 or 19.5 nm channel ER profile is available, or if they do not span an altitude region reasonable to do the above initial value calculations, the exospheric temperature and the O and the N$_2$ number densities from The Naval Research Laboratory Mass Spectrometer and Incoherent Scatter radar 2.0 (NRLMSIS-2.0; hereafter, MSIS) model \mbox{\citep{MSIS}} outputs are used for the initial values, using the F10.7 and Ap values at the time of the SUVI observation. Furthermore, in any case where the initial value algorithm produces an nonphysical initial value (e.g. NaN or negative values caused by ER measurement noise at high altitudes producing poor LOS density inversions), the corresponding MSIS value will be used instead, so that the retrieval can still progress. In all of these cases, there is a data quality flag in the density and temperature data file indicating if MSIS was used and for which specific initial values it was used for.}{The absorption cross sections for O and N$_2$ in the 17.1 nm and 19.5 nm passbands are nearly equal \mbox{\citep{Thiemann2017}}; that is, $\sigma_{O,17.1} \approx \sigma_{N_2,17.1}$ and $\sigma_{O,19.5} \approx \sigma_{N_2,19.5}$. This property is leveraged to derive initial values for the exospheric temperature (T$_{exo}$), atomic oxygen density ($n_O(z_{0,O})$), and molecular nitrogen density ($n_{N_2}(z_{0,N_2})$), by assuming the cross sections are exactly equal and inverting Equation {\ref{eqn:beer_lambert}} to obtain the total line-of-sight (LOS) density of O(z) + N$_2$(z) from the available 17.1 nm and 19.5 nm altitude-averaged extinction ratio (ER) profiles. At high altitudes, where $ER > 0.75$, atomic oxygen dominates the composition and the temperature is approximately isothermal. Under these conditions, an exponential fit to the O(z) + N$_2$(z) LOS density yields an initial estimate of the O-only LOS density. This profile is then inverted using the Abel transform (Equation {\ref{eqn:abel_transform}}) to derive a preliminary estimate of the atomic oxygen number density, $n_O(z)$. The value of this profile at a reference altitude of $z_{0,O} = 300$ km provides the initial value $n_O(z_{0,O})$. Assuming near-isothermal conditions, the slope of a linear fit to $\ln \left[n_O(z)\right]$ at high altitudes corresponds to the inverse scale height of atomic oxygen (per Equations {\ref{eqn:diffusive_equalibrium}} and {\ref{eqn:scale_height}}), from which the initial exospheric temperature, T$_{exo}$, is derived. The residual between the measured total LOS density (O + N$_2$) and the fitted O-only LOS density provides an estimate of the N$_2$ LOS density, which is likewise inverted via the Abel transform to yield $n_{N_2}(z)$. The value at $z_{0,N_2} = 250$ km defines the initial value $n_{N_2}(z_{0,N_2})$.

If neither the 17.1 nm nor 19.5 nm ER profiles are available, or if their vertical coverage is insufficient for performing the above calculations, then initial values for T$_{exo}$, $n_O(z_{0,O})$, and $n_{N_2}(z_{0,N_2})$ are obtained from the Naval Research Laboratory Mass Spectrometer and Incoherent Scatter Radar model (NRLMSIS-2.0; hereafter MSIS) \mbox{\citep{MSIS}}, using the F10.7 and Ap indices corresponding to the time of the SUVI observation. Additionally, if the initial value algorithm produces a nonphysical result (such as NaNs or negative values due to ER noise at high altitudes) MSIS values are used in place of the derived ones to allow the retrieval to proceed. In all such cases, data quality flags are provided in the output density and temperature files to indicate when MSIS values were used and for which initial parameters.}

\begin{figure}[H]
    \centering
    \includegraphics[width=\linewidth]{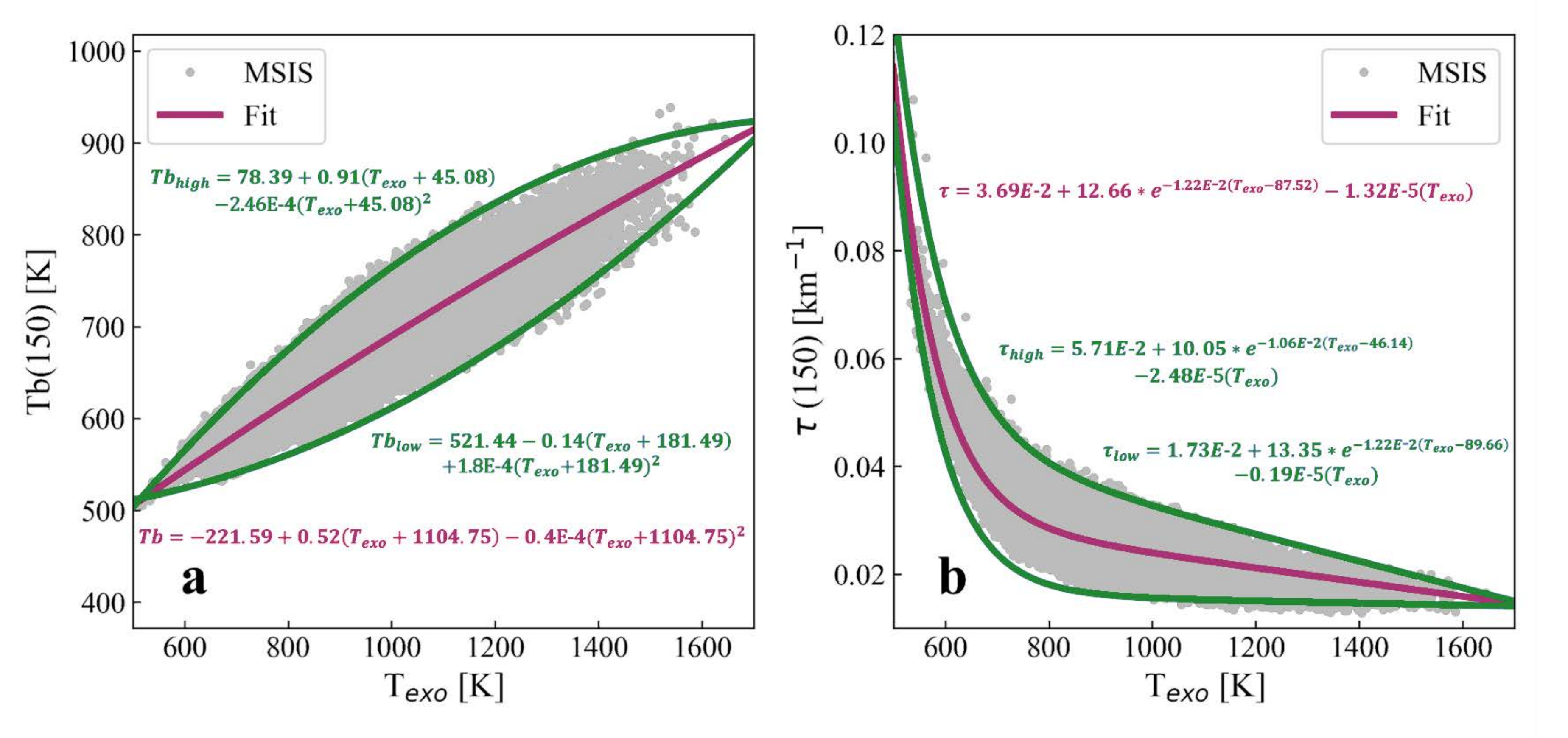} 
    \caption{\textbf{(a)} Estimates of constraints on fitted parameters. Scatter-plot (gray) and fit (purple) of T$_b$ vs. Texo derived from MSIS\remove{-2.0} for the GOES-R/SUVI observing locations and times from 2018-2023, with the allowable envelope of forward model fit values in green. \textbf{(b)} Same as in \textbf{(a)} but for $\tau$ vs. Texo from MSIS\remove{-2.0}.}\label{fig:seed_values}
\end{figure}

\change{Using the initial value of T$_{exo}$ found above, the initial values of T$_b$ and $\tau$ are then estimated using fits between these parameters and the T$_{exo}$ reported by MSIS for all GOES-16 SUVI occultation observing times and locations from 2018-2023, shown in purple in Figure \mbox{\ref{fig:seed_values}}.}{Using the initial value of T$_{exo}$ derived above, initial estimates of T$_b$ and $\tau$ are obtained through empirical fits relating these parameters to the MSIS-reported T$_{exo}$. These fits are based on all GOES-16 SUVI occultation observations from 2018 to 2023 which are shown in purple in Figure \mbox{\ref{fig:seed_values}}.} Furthermore, during the forward model run, T$_b$ and $\tau$ are constrained to the envelopes of expected values shown in green in Figure \ref{fig:seed_values}\add{ for each iteration of the forward model}. The ER profiles of SUVI observations are not strongly sensitive to these shaping parameters, \add{because they primarily alter the temperature profile at the lowest observed altitudes where signal-to-noise in the observed ER is high. Sensitivity studies of the forward model showed that values of $\tau$ and T$_b$ well outside of realistic climatologically, would produce similar O, N$_2$ and T$_{exo}$ retrievals as those derived with realistic $\tau$ and T$_b$ values, for the same occultation.} \change{and thus}{Thus,} MSIS is used as a means to keep these fitted parameters tied to the physical climatology, represented by the green envelopes in Figure {\ref{fig:seed_values}}. The observations are much more sensitive to $T_{exo}$, $n_O(z_0)$, and $n_{N_2}(z_0)$, the initial values of which are typically derived independent of MSIS.
\subsection{Absorption Cross Sections}\label{subsec:CS_Uncertainty}
\change{The}{Inferring densities from the} extinction ratio is highly sensitive to precise knowledge of the effective absorption cross section for both O and N$_2$. Preliminarily, LOS densities were derived by inverting Equation \ref{eqn:beer_lambert} using the SUVI ER profiles and effective total absorption cross sections\add{. These effective total absorption cross sections are} derived by weighting the respective SUVI channel response, from \citet{Darnel2022}, with the shape of the solar spectrum and the total absorption cross section presented in \citet{Fennelly1992} -- hereafter referred to as ``lab'' total effective cross sections. However, when using two different pairs of channel extinction ratios (e.g., 17.1 nm \& 19.5 nm and 17.1 nm \& 30.4 nm) from the same occultation observation to directly solve for the LOS density, the retrieved densities would routinely differ from one another. In theory, these density measurements should be nearly identical, as the LOS column is nearly identical from channel to channel during a given observation. \change{These discrepancies appeared to be systematic between observations, implying a discrepancy in these effective total absorption cross sections that use the lab absorption measurements and the SUVI channel responses rather than of random errors such as altitude registration or detector noise between consecutive channel measurements.}{The discrepancies appeared systematic across observations, pointing to differences in the effective absorption cross sections.}\add{ An initial ad-hoc sensitivity study showed that slight ($<$5\%) adjustments would remove discrepancy in retrievals from different channel pairs for the same occultation.}

\change{To correct for this discrepancy due to the effective cross sections more systematically, LOS densities were constructed using MSIS modeled number densities coinciding with SUVI observations.}{To systematically correct for discrepancies caused by the effective cross sections, LOS densities were constructed using MSIS-modeled number densities corresponding to each SUVI observation.} These LOS densities were then treated as \change{“ground-truth”}{a reference atmosphere} and the absorption cross section were solved for using the SUVI measured extinction ratio as in Equation \ref{eqn:least_squares_cross_section}.
\begin{equation}\label{eqn:least_squares_cross_section}
    \begin{bmatrix}-\ln\left(\frac{ER_{\lambda,z}}{e^{-\sigma_{ O_2,\lambda}N_{O_{2(MSIS)},z}}}\right)\end{bmatrix}=\begin{bmatrix}
        \sigma_{O,\lambda} &
        \sigma_{N_2,\lambda}
    \end{bmatrix}\begin{bmatrix}
        N_{O_{(MSIS)},z}\\ N_{N_{2(MSIS)},z}
    \end{bmatrix}
\end{equation}
\change{Where}{Here} the contribution from the minor species, O$_2$, in the reference atmosphere is \change{factored}{divided} out of the measured ER \add{(from its contribution in Equation \mbox{\ref{eqn:beer_lambert}}}) so that the effective cross sections of O and N$_2$ can be solved for. \add{The total effective channel cross sections for O$_2$ were fitted using the values from \mbox{\cite{Fennelly1992}}, as described for O and N$_2$ above -- determined as $6.88\pm0.97$, $8.46\pm1.26$, and $11.77\pm1.28$ [$\times10^{-18}$ cm$^2$] for the 17.1, 19.5 and 30.4 nm channels, respectively}. Using equation \ref{eqn:least_squares_cross_section}, the total effective absorption cross sections for O and N$_2$ were derived for every SUVI occultation observation from the September equinox eclipse season of 2018 through the March equinox eclipse season of 2024\add{, where at least two distinct channel ER measurements are present}. When taking the ratio of these derived cross sections to the effective lab cross section for the SUVI 19.5 nm passband, a nearly constant \change{offset}{factor}\change{ over this time period -- spanning a wide range of solar activity and observational latitudes -- was found}{ was found}. \add{This result implies a small  (within the reported cross section uncertainties) adjustment to the cross sections is needed to make the density retrievals self consistent when using different channel pairs.} Table \ref{table:cs_percent_diff} summarizes the adjusted cross sections that are found when applying this ratio to the lab values\remove{, which are used in the SUVI forward model retrievals,} along with percent differences between the two. 

\begin{table}[H]
\begin{center}
\caption{Lab total effective absorption cross sections compared to the derived adjusted cross sections ratios and absolute adjusted cross section which are used for SUVI density retrievals.}\label{table:cs_percent_diff}
\makebox[\textwidth][c]{%
\begin{tabular}{c c c c c c c}
\toprule
   \multicolumn{1}{c}{}&\multicolumn{3}{c}{O Cross Section} & \multicolumn{3}{c}{N$_2$ Cross Section}\\
    & 17.1 nm & 19.5 nm & 30.4 nm & 17.1 nm & 19.5 nm & 30.4 nm \\
    \cmidrule(lr){2-4} \cmidrule(lr){5-7}
    Lab Total Effective [$\times10^{-18}$ cm$^{2}$] & 3.37$\pm0.11$ & 4.23$\pm0.08$ & 7.67$\pm0.28$ & 4.51$\pm0.10$ & 5.85$\pm0.08$ & 11.77$\pm0.31$ \\
    Ratio [$\sigma_{s,\lambda}/\sigma_{s,195}$] & 0.79$\pm$0.05 & 1 & 1.79$\pm$0.08 & 0.79$\pm$0.07 & 1 & 2.10$\pm$ 0.16\\
    Adjusted  [$\times10^{-18}$ cm$^{2}$] & 3.35$\pm$0.22 & 4.23$\pm$0.08 & 7.60$\pm$0.37 & 4.68$\pm$0.41& 5.85$\pm$0.08 & 12.39$\pm$0.95\\
    Diff. [\%] & -0.59 & 0.00 & -0.91 &3.77 & 0.00 & 5.27\\
\bottomrule
\end{tabular}%
}
\end{center}
\end{table}

\add{In theory, any lab cross section of the three channels could have been used to ground the adjustment ratios to the lab values.} \add{If instead the 17.1 or 30.4 nm lab cross sections were used as the grounding reference value, the retrieved the O and N$_2$ number densities would change as a constant percent difference equal to the difference between adjusted cross sections for each species. Specifically, using the 17.1 nm lab cross sections would result in O densities being lower by 0.76\% and N$_2$ densities higher by 3.81\%; using the 30.4 nm lab cross sections would lead to offsets of -1.35\% for O and +4.80\% for N$_2$. To validate this, the forward model was run using both 17.1 nm and 30.4 nm lab cross section values as the reference values, and the resulting O and N$_2$ density outputs were compared to the nominal SUVI occultation data pipeline outputs to confirm that these offsets were as expected. During this validation, the percent difference offsets were found to be constant and matched the expected differences above. Furthermore, no significant percent change was found in the derived exospheric temperature between any choice of reference lab cross section reference. }

\change{It is important to note that while this adjustment used MSIS density profiles as reference atmospheres, because the adjustment is taken as ratio to the lab 19.5 nm effective cross sections of O and N$_2$, the magnitude of the measured densities remains independent of MSIS.}{It is important to emphasize that although the adjustment process uses MSIS density profiles as reference atmospheres, the resulting measured densities remain independent of MSIS. This independence arises because the adjustment is applied as a ratio relative to the laboratory-derived 19.5 nm effective cross sections for O and N$_2$ preventing the correction from tying the SUVI absolute density values to that of the MSIS climatology. The adjusted cross sections remain within the measurement uncertainty of the original lab values. These adjustments are most likely attributable to uncertainty in the total absorption cross section, as the derived adjustment ratios remain nearly constant across varying solar conditions, altitude ranges, and ER channel sampling. This consistency suggests that systematic errors in instrument response or altitude registration are unlikely to be the primary cause. However, the adjustment could reflect a systematic bias stemming from the omission of O$_2$ absorption in the inversion process used to calculate the adjustment ratios. This potential bias is likely mitigated by the subtraction of the O$_2$ contribution -- estimated from MSIS densities -- from the SUVI extinction signal prior to retrieval. Furthermore, even if the MSIS estimates deviate from realistic densities, the impact of O$_2$ is further reduced because the ratio adjustment fits are performed across all altitudes of the direct inversion, most of which are above where O$_2$ contributes meaningfully to the absorption.}

\begin{figure}[H]
    \centering
    \includegraphics[width=.95\linewidth]{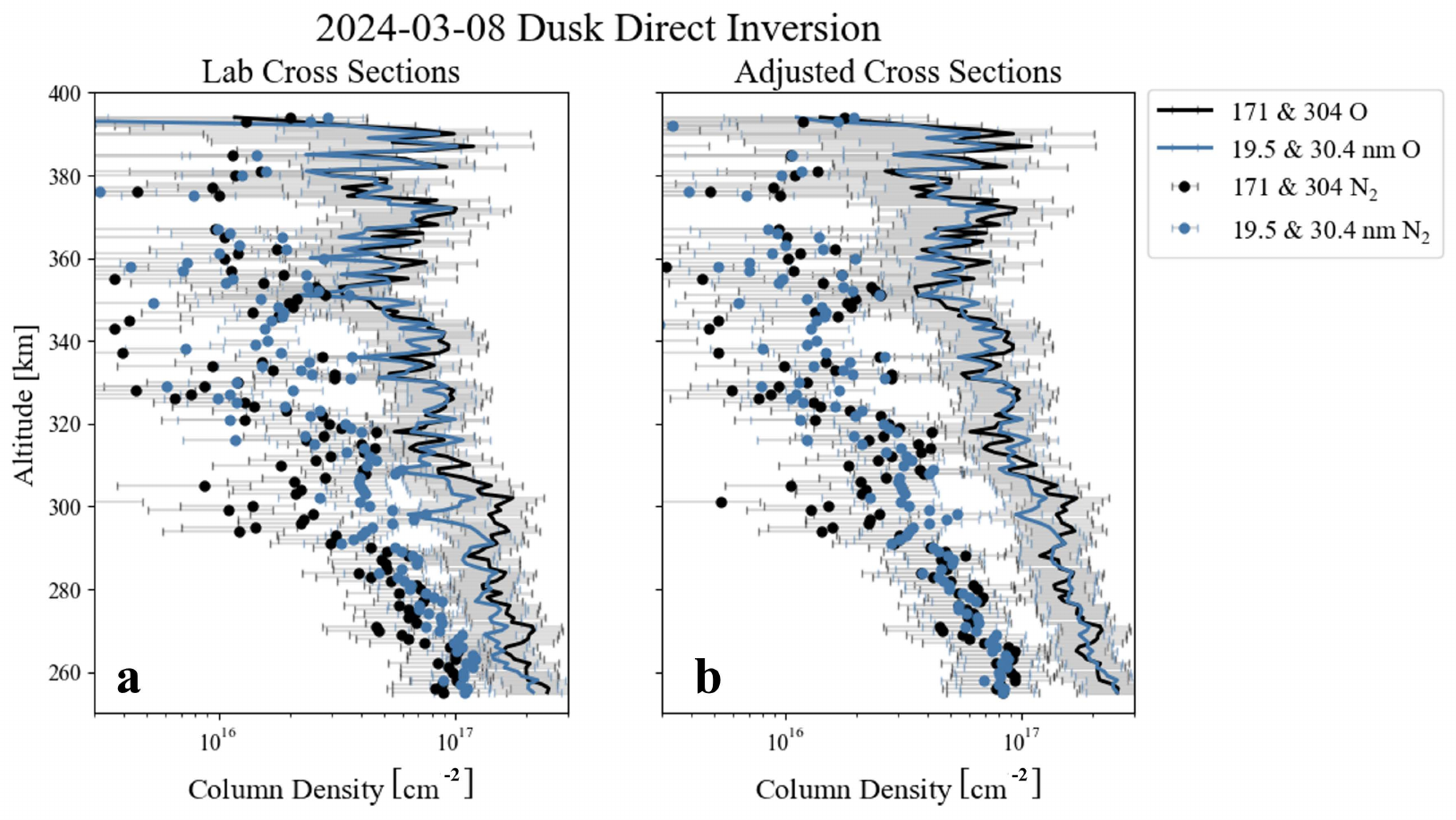} 
    \caption{\textbf{(a)} Direct inversion of Equation \ref{eqn:beer_lambert}, using experimentally determined O and N$_2$ total absorption cross sections from \citet{Fennelly1992}, to derive O and N$_2$ LOS densities from GOES-16/SUVI 2024-03-08 Dusk ER observations. \textbf{(b)} Same as in \textbf{(a)} but using the derived adjusted SUVI O and N$_2$ cross sections. \add{1-$\sigma$ uncertainty end caps are derived from the average ER profiles uncertainties propagated through the direct inversion.}}\label{fig:direct_cs}
\end{figure}

\remove{Figure {\ref{fig:direct_cs}} shows an example direct density retrieval from the 2024-02-29 dawn SUVI occultation produced by simply inverting Equation {\ref{eqn:beer_lambert}} using the lab total absorption cross section values compared to the retrieval using the derived adjusted cross sections.}

\remove{From Figure {\ref{fig:direct_cs}}, the moderated adjustments to the 17.1 nm and 30.4 nm effective cross sections drastically improve the retrieval quality, particularly for N$_2$. It is also important to note, that while this adjustment used MSIS density profiles as reference atmospheres, because the adjustment is taken as ratio to the lab 19.5 nm effective cross sections of O and N$_2$, the magnitude of the measured densities remain independent of MSIS.} 

\add{Error in the effective absorption cross sections will result in inconsitent retrievals when ERs from different channel pairs are directly inverted for density. To evaluate the adjusted cross sections, both the laboratory and adjusted values were applied in the direct inversion of Equation 1 to derive O and N$_2$ LOS densities. This analysis was conducted across all available SUVI observations, using both the 17.1 \& 30.4 nm and 19.5 \& 30.4 nm channel pairs. Figure {\ref{fig:direct_cs}}, shows an example of this calculation comparing the direct inversion when using the lab cross sections and the adjusted cross sections. As shown in {\ref{fig:direct_cs}}b, the retrieval quality improves when using the adjusted cross sections, as evidenced by a reduction in the variance between O and N$_2$ densities derived from different channel pairs. This indicates that the extinction ratio (ER) measurements are more self-consistent for this observation. When comparing densities derived from the 17.1 \& 30.4 nm pair to those from the 19.5 \& 30.4 nm pair, the standard deviation between the two was reduced by 25\% for O densities and 46\% for N$_2$ densities when using the adjusted cross sections, relative to lab weighted cross sections, across all observations. Furthermore, O and N$_2$ densities and thermospheric temperature were derived using the forward model for all SUVI observations using both the lab and adjusted cross sections. The mean difference between the derived values using the lab cross sections verses those derived using the adjusted cross sections was -2\% and +3\% for the O and N$_2$ number densities at 250 km, respectively, and +1\% for exospheric temperature. Overall, this small adjustment to the magnitude of the cross sections is within the reported measurement error of the lab effective cross sections for all adjustments, greatly improve the self-consistency of SUVI direct inversion measurements and only result in minor changes to the mean retrieved thermospheric parameters in the forward model. As such, these adjusted effective cross sections are used for all SUVI occultation forward model data processing, rather than the effective cross sections derived using the lab total absorption cross sections.}

\section{Error and Uncertainty}\label{sec:error_fwd_model}
\begin{figure}[H]
    \centering
    \includegraphics[width=.95\linewidth]{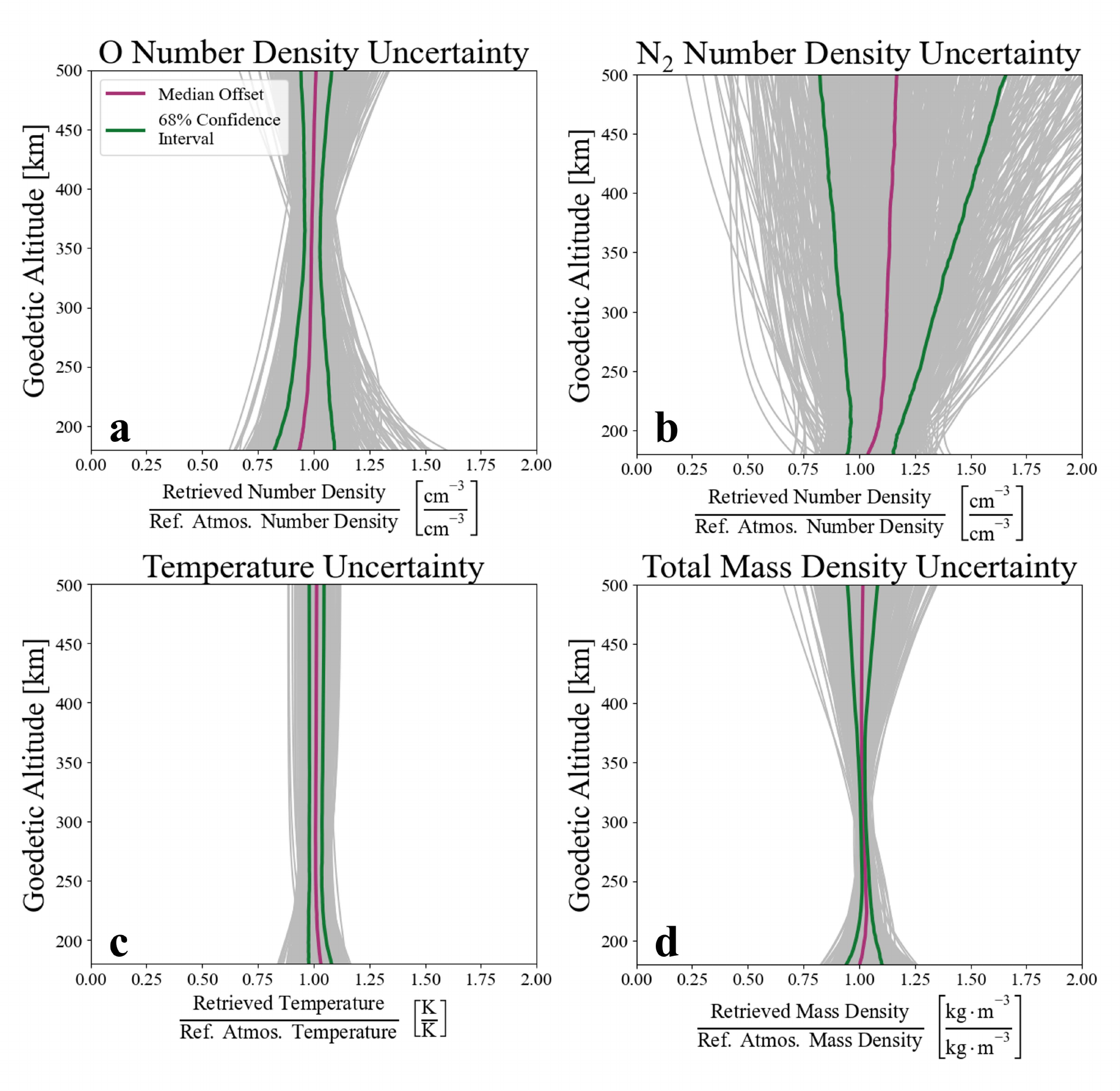} 
    \caption{\textbf{(a)} Ratio of retrieved \change{O LOS density to ground truth LOS density}{O number density to reference atmospheric density}. Gray profiles are ratios from all model runs, purple is the \change{mean ratio offset, green is 1-$\sigma$ standard deviation around the mean}{median ratio offset, green is the 68\% confidence interval around the median ratio}. \textbf{(b)} Same as in \textbf{(a)} but for N$_2$ \change{LOS density}{number density}. \textbf{(c)} Same as in \textbf{(a)} but for neutral temperature. \textbf{(d)} Same as in \textbf{(a)} but for total mass density.}\label{fig:uncertainty}
\end{figure}

\remove{The uncertainty of the forward modeled atmospheres was characterized by deriving synthetic ERs from MSIS derived LOS densities for all SUVI occultation observations, as described in Section {\ref{subsec:CS_Uncertainty}}. The MSIS atmosphere was treated as "ground-truth" and the generated synthetic ERs were then restricted to the same SUVI observation altitudes for each channel observation and superimposed with the real 1-$\sigma$ observational ER noise derived from ER image altitude binning. These synthetic ER profiles were then passed through the forward model described in Section {\ref{subsec:forward_model}}. Figure {\ref{fig:uncertainty}} shows the ratio of the retrieved atmospheric parameters to those for the ground-truth atmosphere.} 
\change{The uncertainty of the forward model retrieval method was characterized through a Monte Carlo approach. For each SUVI observation, synthetic MSIS reference atmospheres consisting of O, N$_2$ and O$_2$ were modeled using the same location, time, solar and geomagnetic conditions during the observation. These are then used to derive synthetic ER curves over the SUVI channel passbands, using Equation {\ref{eqn:beer_lambert}}, to include absorption from all three thermospheric species. While O$_2$ is a minor species over most observing altitudes of the SUVI occultations, at the lowest altitudes the optical depth can become significant compared to that of atomic oxygen (e.g. $\nicefrac{\tau_{\text{O}_2}}{\tau_\text{O}}\sim0.15$ around 200 km in some climatological conditions).  These ERs were limited to the real observation altitudes and channel coverage of the SUVI observation, and were superimposed with the SUVI ER measurement noise for each given observation. Initial forward model parameter values were derived from these ERs (as in Section {\ref{sec:seed_values}}), excluding cases when the initial values default to the MSIS values, and passed through the forward model (Section {\ref{subsec:forward_model}}). The resultant retrieved temperature and O and N$_2$ densities were then compared to the input reference atmosphere in order to characterize the systematic and random uncertainty imposed by the forward model retrieval process and ER measurement noise and sampling, as well as any systematic error imposed by ignoring O$_2$ in the model.}{The uncertainty of the forward model retrieval method was quantified using a Monte Carlo approach. For each SUVI observation, synthetic MSIS reference atmospheres were generated -- including O, N$_2$, and O$_2$ -- using the same geographic location, time, and solar and geomagnetic conditions as the observation. These reference atmospheres were used to compute synthetic ER curves across the SUVI channel passbands via Equation 1, incorporating absorption from O, N$_2$, and O$_2$. Although O$_2$ is a minor constituent at most altitudes probed by SUVI occultations, its contribution to the optical depth can become non-negligible at lower altitudes (e.g. $\nicefrac{\tau_{\text{O}_2}}{\tau_\text{O}}$ can be on the order of 15\% at 200 km, in some climatological conditions). The synthetic ERs derived from these reference atmospheres were constrained to the same altitude range and channel coverage as the SUVI observations, with random noise added to match the measurement uncertainty in each case.

Initial forward model parameter values were derived from these synthetic ERs (following the procedure in \mbox{\ref{sec:seed_values}}), excluding cases where the initial values defaulted directly to MSIS. These parameters were then input into the forward model (Section \mbox{\ref{subsec:forward_model}}), and the resulting retrieved temperatures and O and N$_2$ densities were compared to the reference atmosphere to characterize both systematic and random uncertainties introduced by the forward model, ER noise, and the exclusion of O$_2$ in the retrieval.} Figure \mbox{\ref{fig:uncertainty}} shows the ratio of the retrieved atmospheric parameters to those for the reference input atmosphere, for all model runs for observations from the September equinox eclipse season of 2018 through the March equinox eclipse season of 2024.
\change{This analysis provides a systematic offset of each retrieved parameter, shown in purple in Figure {\ref{fig:uncertainty}}, which, for an example very low earth orbit (VLEO) altitude of 250 km, were found to be -2\%, +11\%, +1\% and 3\% for O number density, N$_2$ number density, neutral temperature, and total mass density, respectively.}{This analysis provides a systematic offset estimate for each retrieved parameter, illustrated in purple in Figure~{\ref{fig:uncertainty}}. At a representative very low Earth orbit (VLEO) altitude of 250~km, the systematic offsets were found to be approximately -2\% for O number density, +11\% for N$_2$ number density, +1\% for neutral temperature, and +3\% for total mass density.} These average offsets are applied as standard corrections, with respect to altitude, to the reported density and temperature profiles in the SUVI occultation data product\add{ to correct for these systematic biases}. \change{The random uncertainties are then calculated as the 68\% confidence interval around the systematic offset, which for a tangent height of 250 km were found to be -7/+8\%, -17/+16\%, -3/+3\% and -2/+2\% for O number density, N$_2$ number density, neutral temperature and mass density, respectively.}{The random uncertainties are then calculated as the 68\% confidence intervals around the systematic offset, which, for a tangent height of 250~km, were found to be approximately -7/+8\% for O number density, -17/+16\% for N$_2$ number density, -3/+3\% for neutral temperature, and -2/+2\% for total mass density.} These calculated systematic offsets and standard error for all LOS and number densities as well as temperature profile are reported in the SUVI occultation data files.

\add{The relatively higher systematic bias and retrieval variability of N$_2$ (Figure \mbox{\ref{fig:uncertainty}}b) compared to that of O (Figure \mbox{\ref{fig:uncertainty}}a), is potentially due to the fact that the density of N$_2$ is much less than that of O, except at the very lowest altitudes that SUVI measures, meaning absorption due to O contributes to the bulk of the signal where the signal-to-noise is the highest and is more easily accounted for by the forward model. This also explains why the individual N$_2$ profile error generally increases with altitude where the contribution to the extinction becomes less significant. 

In addition, at lower altitudes absorption by O$_2$ may become non-negligible. Since O$_2$ is excluded from the forward model, the retrieval may compensate by overestimating N$_2$, contributing to the observed systematic bias. To explore this, the uncertainty analysis was repeated using synthetic atmospheres with O$_2$ densities set to double and half those in MSIS. In the doubled-O$_2$ case, the systematic offset of N$_2$ increased by 11\% and total mass density increased by 5\% (shifting the entire confidence interval envelopes) at a tangent altitude of 250~km, while other parameters remained largely unaffected. In the halved-O$_2$ case, the systematic offset for N$_2$ decreased by 3\% and mass density decreased by 2\% at a tangent altitude, with no significant change was found for any of the other thermospheric parameters. This suggests that the retrieval is largely insensitive to O$_2$ near MSIS-reported levels but may show an enhanced N$_2$ bias if O$_2$ is significantly increased. However, recent GOLD observations \mbox{\citep{Greer2022, Greer2025}} indicate that MSIS 2.0 often overestimates O$_2$, especially near dawn and during solar minimum. This further suggests that O$_2$ is unlikely a dominant source of error in these retrievals.} 

The LOS density systematic offsets and standard error were found to be within $<1\%$ for all altitudes as those reported for both O and N$_2$ above. This is expected because of the direct relationship between LOS density and number density with assumed spherical symmetry. \change{Without assuming spherical symmetry around the terminator plane these uncertainties may be slightly different, but these differences should be small ($<$1\%) as the bulk of the absorption is at the terminator plane due to the exponentially decreasing nature of the thermospheric densities with altitude.}{Without assuming spherical symmetry around the terminator plane, these uncertainties may be slightly different. However, previous studies on ozone solar occultations at Mars have shown that the effects of spherical symmetry depend on the magnitude of the species' gradient \mbox{\citep{Mars}}. Ozone, which is highly photochemically active, has been observed to vary by up to two orders of magnitude across the terminator, leading to a 20\% bias in retrieved densities when assuming spherical symmetry. However, when the ozone gradient was more gradual — varying by tens of percent — the assumption of spherical symmetry had a negligible effect (less than 1\%). For all SUVI observing passbands, approximately 80\% of the extinction occurs within $\pm$500 km across the terminator. Over these distances, the thermospheric O and N$_2$ gradients are also on the order of tens of percent, suggesting that the impact of the spherical symmetry assumption on the retrieved densities should remain small.} 

\add{Additional sources of error may arise in the case when an initial value cannot be determined from the ER profiles themselves and the MSIS values are used instead. To characterize this uncertainty, the forward model was run using the MSIS values for all initial parameters for all observations from the September 2018 occultation season through the September 2024 season. When comparing the derived densities and temperatures using MSIS initial values to all forward model runs during the same time period with initial values derived from the ER profiles, the standard error between the two was 3\% and 10\% for the O and N$_2$ number densities at 250 km, respectively, and 1\% for exospheric temperature. As such, the uncertainty when using MSIS for the initial values rather than the ER profiles falls within the overall uncertainty of forward model itself. However, these instances are still reported in the data quality flags for the SUVI occultation data product to inform data users what initial values used MSIS parameters for a given observation.}

\begin{figure}[H]
    \centering
    \includegraphics[width=.95\linewidth]{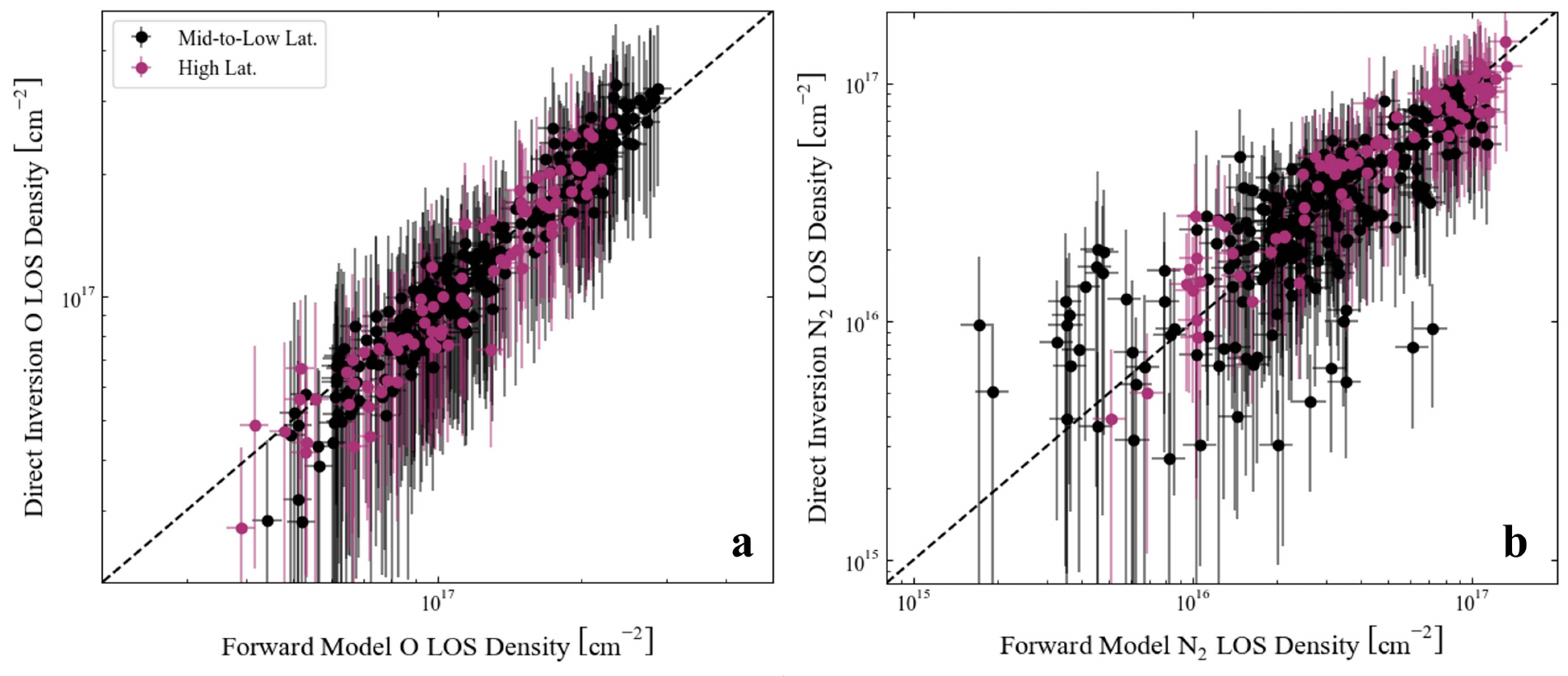} 
    \caption{\textbf{(a)} Comparison of direct inversion and forward modeled LOS density retrievals of O and high latitudes ($\left|\mathrm{Lat}\right|$ $>$ 60\textdegree; purple) and middle to low latitudes (black) at a tangent altitude of 250 km. The linear 1-to-1 agreement between the direct inversion and the forward model is shown as a dashed black line. \textbf{(b)} Same as \textbf{(a)} but for N$_2$ LOS densities.\add{Forward model error bars represent the random uncertainty estimates from Section~{\ref{sec:error_fwd_model}} and direct inversion error bars come from the average ER profile uncertainties propagated through the direct inversion.}}\label{fig:direct_to_forward}
\end{figure}

\add{Lastly, because the forward model assumes the thermosphere to be in a state of diffusive, equilibrium there may be additional error not accounted for in this analysis when this is not the case. This type of error would be the most impactful for polar observations during geomagnetic storm times, when Joule heating and particle precipitation would cause expansion of the neutral atmosphere and diverging winds which would cause deviations from this diffusive equilibrium assumption. To test the impact of this, O and N$_2$ LOS densities were derived, at a tangent height of 250 km, for all SUVI observations from the September 2018 occultation season through the September 2024 season using the direct inversion method, using the same methods as discussed in {\ref{subsec:CS_Uncertainty}}. Differences were then taken between these LOS densities and those derived through the forward model for the same observations. For these differences a two-tail z-test did not reveal a statistically significant difference between the measurement differences at high latitudes ($\left|\mathrm{Lat}\right|\ >\ 60\textdegree$) and those at middle to low latitudes, at the 5\% significance level (p-value $=0.55>0.05$ and p-value $=0.52>0.05$ for O and N$_2$ LOS density differences between forward model and direct inversion measurements, respectively). Additional filtering of observations by geomagnetically active times (Ap $>$ 32) and high latitude also did not reveal a statistical difference in the measurement differences when compared to those at middle to low latitudes (p-value $=0.49>0.05$ and p-value $=0.60>0.05$ for O and N$_2$ LOS density differences between forward model and direct inversion measurements, respectively). Because the LOS densities derived with the direct inversion (which makes no assumptions of the thermospheric state) show good agreement with the forward model, any error due to the assumption diffusive equilibrium in the forward model is likely captured in the uncertainty of the forward model measurement even in cases where we expect the diffusive equilibrium assumption to be violated. This agreement is highlighted by Figure {\ref{fig:direct_to_forward}} where, while there is variance especially for the N$_2$ direct inversion densities due to low signal to noise, overall both the O and N$_2$ retrieved LOS densities agree within 5\% for O and 15\% for N$_2$, which is within the the measurement uncertainty.}
\subsection{Cross Section Uncertainty}\label{sec:cs_error}
\begin{figure}[H]
    \centering
    \includegraphics[width=.95\linewidth]{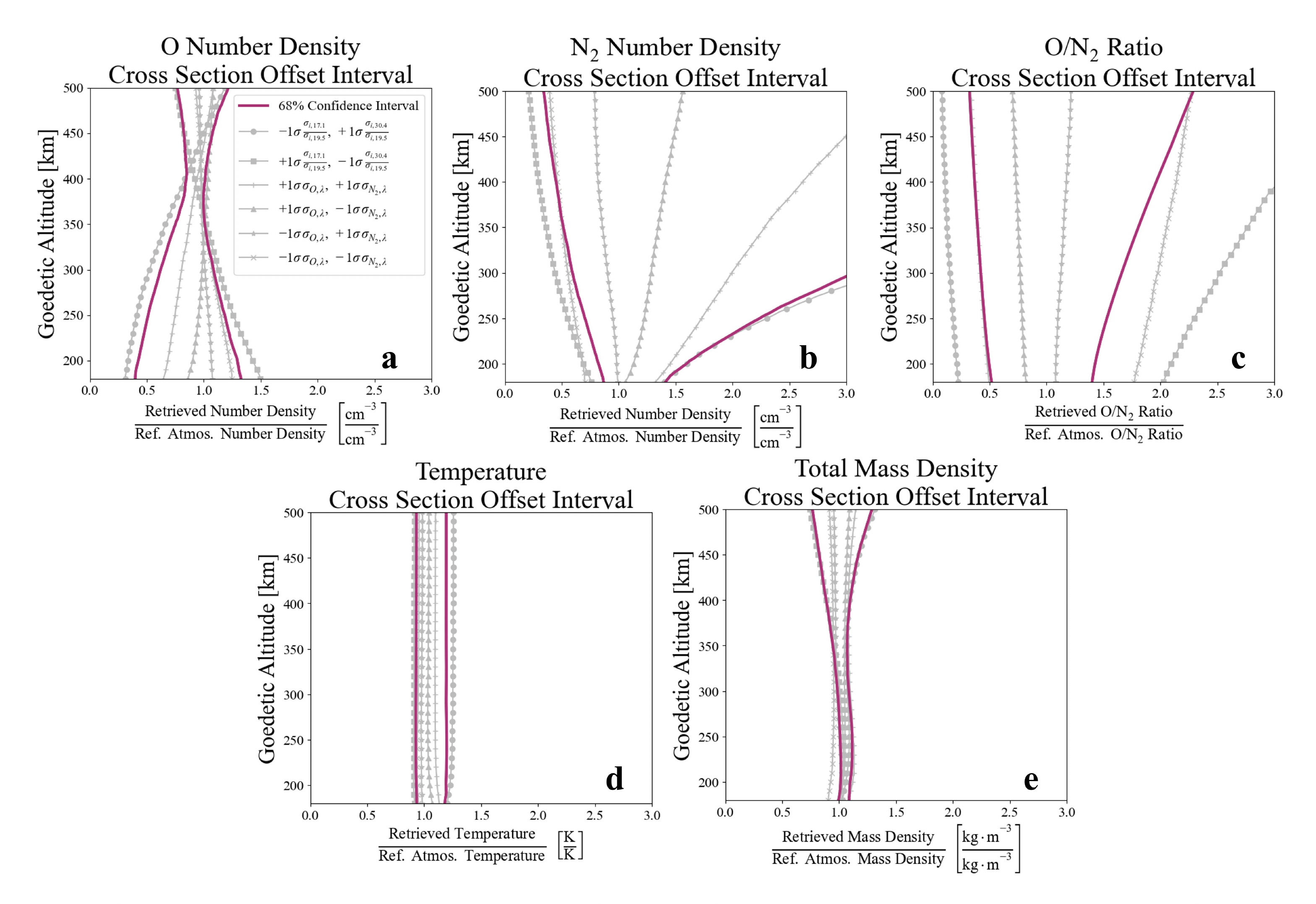} 
    \caption{\textbf{(a)} Ratio bias confidence interval (purple envelope) for retrieved O number density relative to a reference atmosphere, based on Monte Carlo variations in effective cross sections. Gray profiles show specific systematic bias scenarios: gray-circle and gray-square represent opposing 1$\sigma$ shifts in the 17.1/19.5 and 30.4/19.5 nm ratios (see Table~\ref{table:cs_percent_diff}). Other profiles reflect uniform 1$\sigma$ changes across all passbands: gray-plus increases both O and N$_2$ cross sections, gray-triangle increases O and decreases N$_2$, gray-star decreases O and increases N$_2$, and gray-x decreases both. \textbf{(b–e)} Same as \textbf{(a)} but for N$_2$ number density, O/N$_2$ ratio, neutral temperature, and total mass density, respectively.}\label{fig:cs_offsets}
\end{figure}

\add{As previously discussed in Section~{\ref{subsec:CS_Uncertainty}}, SUVI density retrievals are highly sensitive to the total effective cross sections of both O and N$_2$ across the SUVI passbands. Although the channel cross-section ratios derived in Section~{\ref{subsec:CS_Uncertainty}} were intended to mitigate discrepancies observed in the direct inversion of ER profiles, the true effective cross sections may still lie within the uncertainty bounds of those ratios, as well as the uncertainties in the 19.5~nm reference cross sections reported in Table~{\ref{table:cs_percent_diff}}. Any deviation between the true effective cross sections and those adjusted for retrieval would introduce an additional systematic bias across all derived quantities. To evaluate the impact of such potential discrepancies, the analysis presented in Section~{\ref{fig:uncertainty}} was repeated. In this case, synthetic ``ground truth'' ER profiles were generated using perturbed effective cross sections of O, N$_2$ and O$_2$ within their stated uncertainties, while the density retrievals were performed using the adjusted cross sections of O and N$_2$ reported in Section~{\ref{subsec:CS_Uncertainty}}, Table~{\ref{table:cs_percent_diff}}. Figure~{\ref{fig:cs_offsets}} shows how systematic uncertainties in the effective cross sections affect the retrieved O and N$_2$ number densities, neutral temperature, and total mass density profiles. The purple-envelope represents the 68\% confidence interval for the systematic bias introduced by these cross-section uncertainties. Additionally, the figure includes several cross section perturbation scenarios, such as uniform increases or decreases in all cross sections and variations in the slope of effective cross sections across the SUVI passbands. These cases illustrate how such uncertainties bias the retrieved profiles as a function of altitude. At 250~km, the 68\% confidence intervals for the retrieved biases are –49\% to +18\% for O density, –30\% to +124\% for N$_2$ density, –54\% to +53\% for the O/N$_2$ ratio, –7\% to +20\% for temperature, and –1\% to +12\% for total mass density. Although the range of potential biases stems from uncertainties in the effective cross sections, it is important to emphasize that these biases are systematic in nature. That is, for a given set of true effective cross section values, the resulting bias would be consistently applied across all retrievals with a fixed altitude profile within this range, like the gray profiles in Figure~{\ref{fig:cs_offsets}}. This analysis also highlights, that while the retrieval of precise densities is highly sensitive to the accuracy of the cross sections, these uncertainties are largest where each species is a more minor fraction to the total absorption. Because of this, both temperature and total mass density bias uncertainties remain small despite these uncertainties in the cross sections. Furthermore, the biases in individual species densities retreivals act in such a way that the composition ratio biases are relatively constant with altitude for most cross section perturbation scenarios (see gray-profiles in Figure~{\ref{fig:cs_offsets}c}).}

\section{Results}
\subsection{Climatology}\label{sec:climate}
\begin{figure}[H]
    \centering
    \includegraphics[width=.95\linewidth]{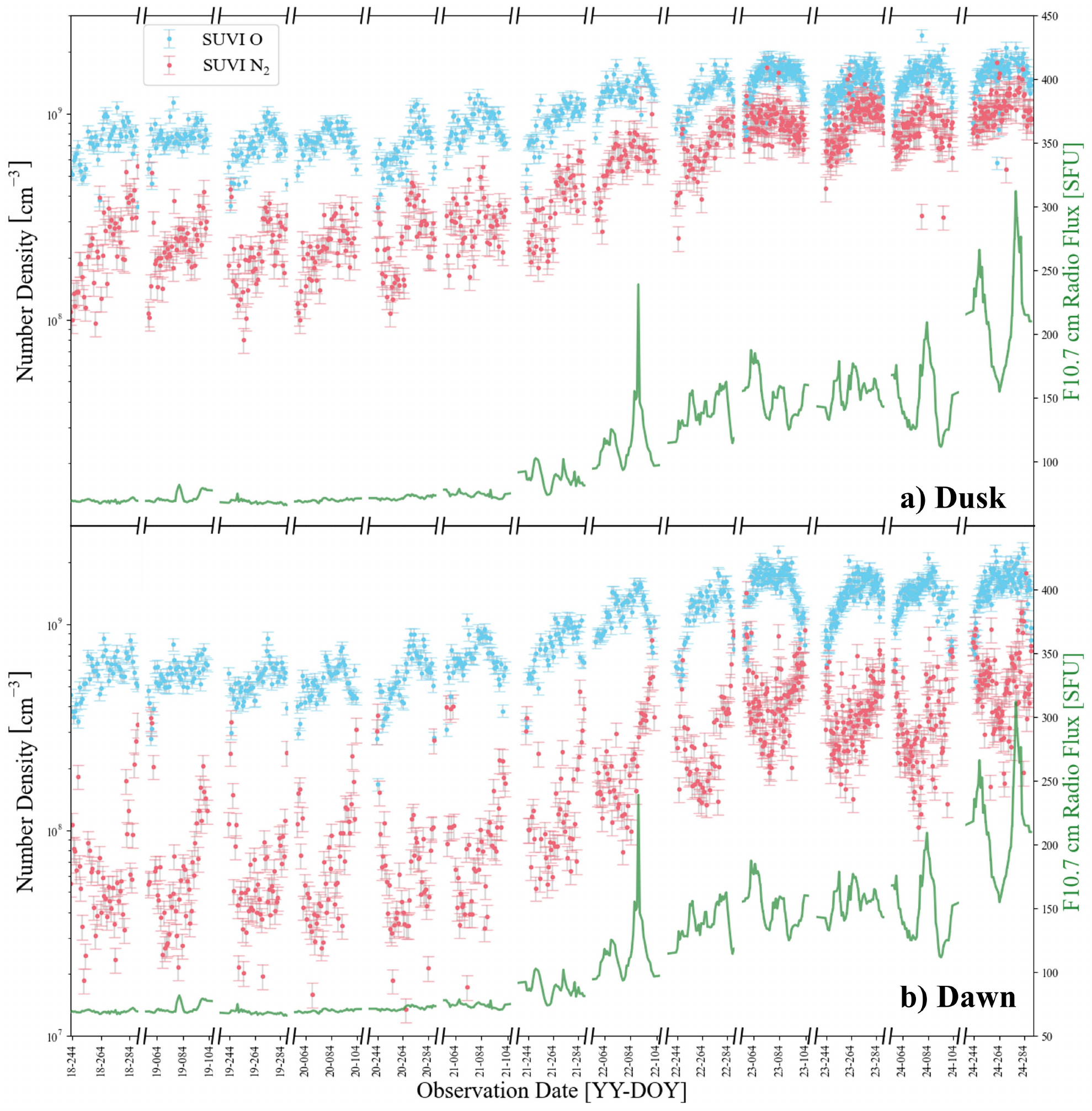} 
    \caption{\textbf{(a)} Dusk observations of O (blue-dots) and N$_2$ (\change{orange}{red}-dots) \change{LOS}{number} density at 250 km over existing GOES-16/SUVI occultation data record. Shown in green is the F$_{10.7}$ radio flux at 1 AU during observing times \citep{Matzka2021-fk}, showing the data coverage from solar minimum leading to the start of Solar Cycle 25 and the current solar activity rise for Solar Cycle 25. \change{\textbf{b}}{\textbf{(b)}} Same as \textbf{(a)} but for dawn-side measurements. \add{All error bars represent the random uncertainty estimates from Section~{\ref{sec:error_fwd_model}}}.}\label{fig:climate_composition}
\end{figure}

\begin{figure}[H]
    \centering
    \includegraphics[width=.95\linewidth]{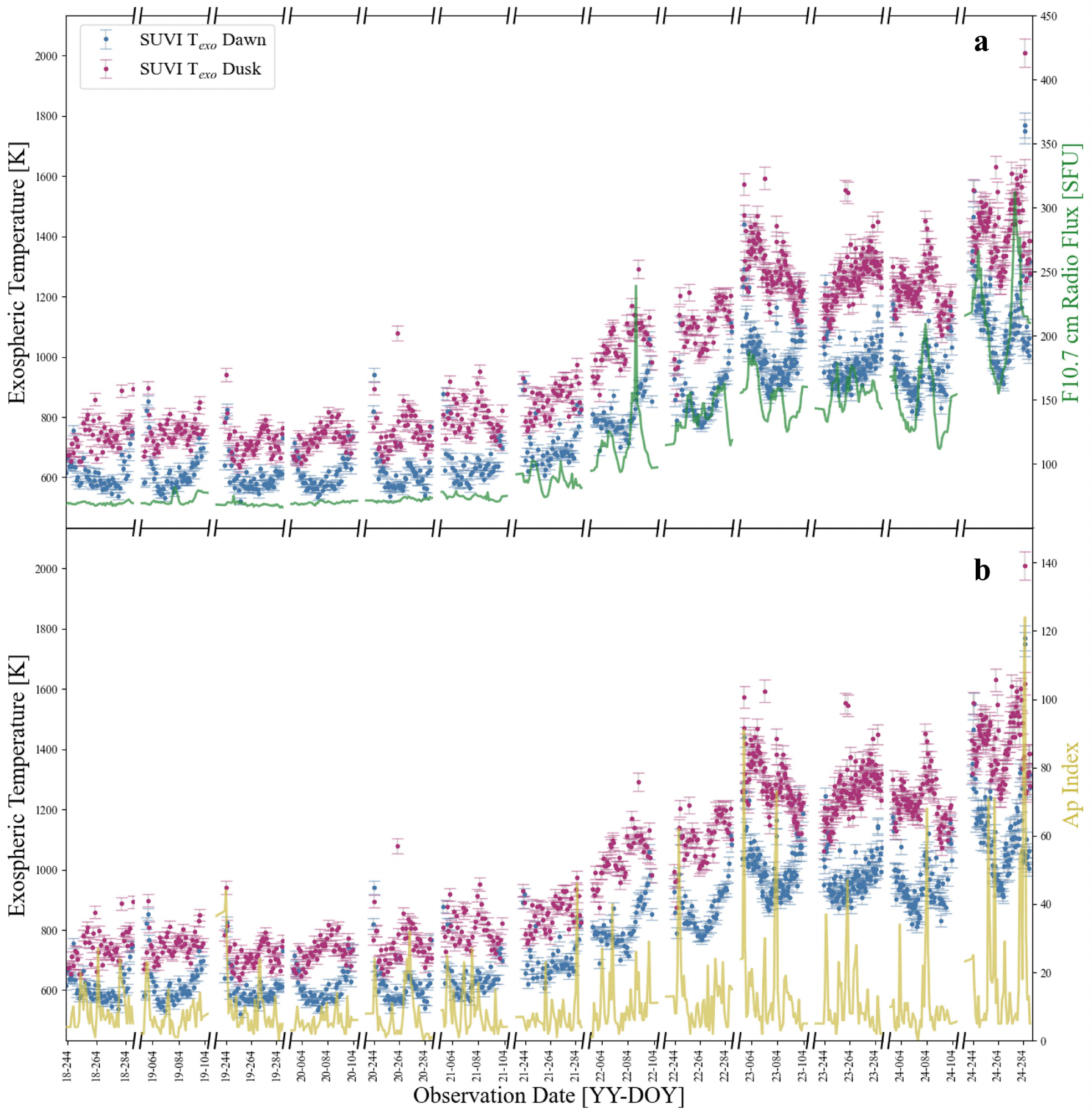} 
    \caption{\textbf{(a)} Similar to Figure \ref{fig:climate_composition} but for exospheric temperature measurements on the dawn (blue-dots) and dusk (\change{red}{purple}-dots) terminator measurements, with the corresponding F$_{10.7}$ radio flux during the observing times in green.\add{\textbf{(b)} Same as \textbf{(a)} but  co-plotted with geomagnetic Ap index in yellow.}\add{ All error bars represent the random uncertainty estimates from Section~{\ref{sec:error_fwd_model}}}.}\label{fig:climate_temp}
\end{figure}

\begin{figure}[H]
    \centering
    \includegraphics[width=.95\linewidth]{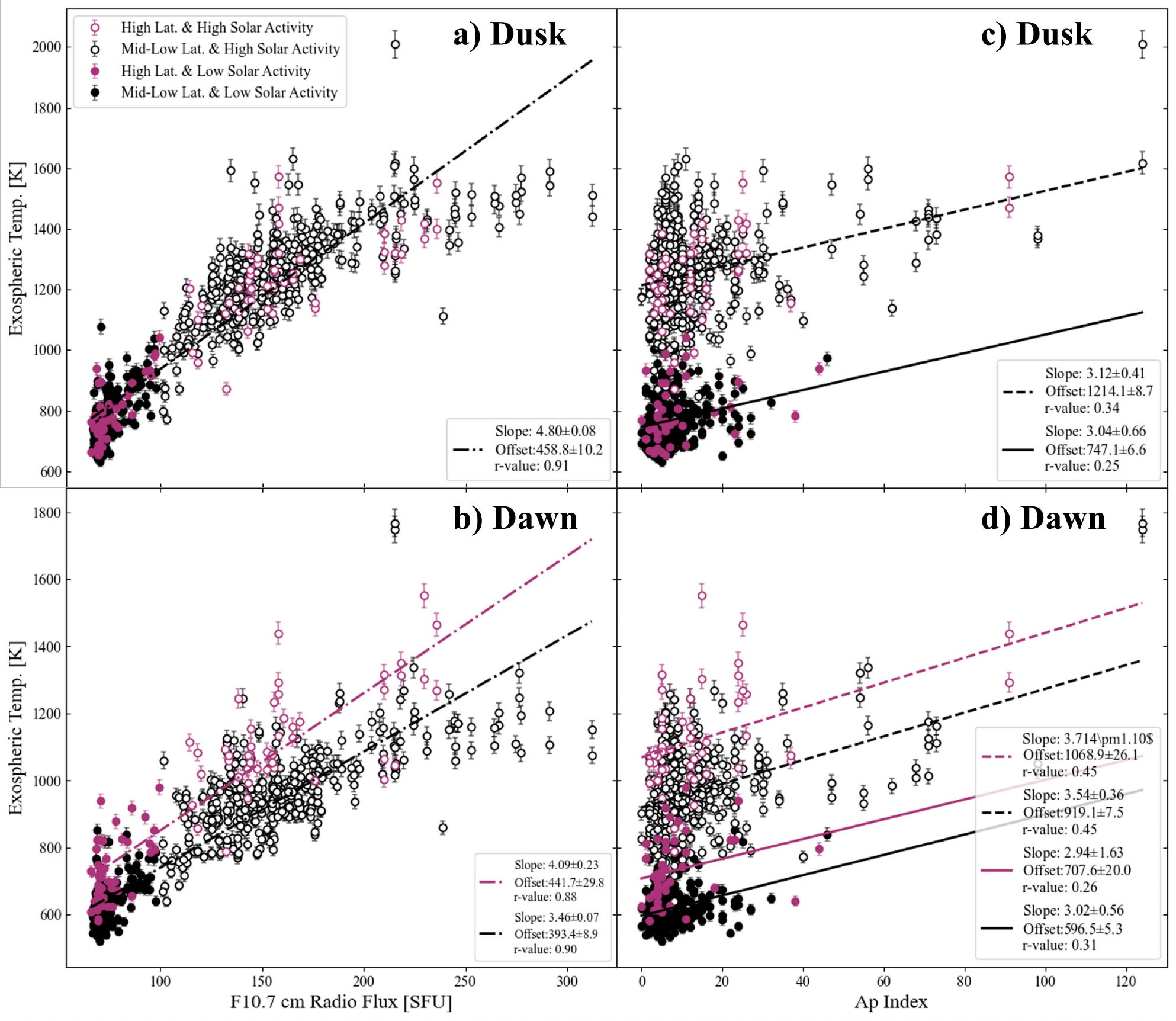} 
    \caption{\textbf{(a)} Exospheric temperature from dusk observations as a function F$_{10.7}$ cm radio flux, with regression line for all data. \textbf{(b)} Same as in \textbf{(a)} but for dawn observations with linear regression performed for both high-latitude ($\left|\mathrm{Lat}\right|$ $>$ 60) observations and mid-to-low ($\left|\mathrm{Lat}\right|$ $\leq$ 60) latitude observations, as high latitudes and mid-to-low latitude distribution differences are statistically different. \textbf{(c)} Exospheric temperature from dusk observations as a function of geomagnetic Ap index, with regression lines both for observations taken during high solar activity (F$_{10.7}\geq100$) and low solar activity (F$_{10.7}<100$). \textbf{(d)} Same as in \textbf{(c)} but for dawn observations, with regression lines given for observations taken at high solar activity and high latitudes, high solar activity and mid-to-low latitudes, low solar activity and high latitudes, and low solar activity and mid-to-low latitudes (all as defined above). \add{All error bars represent the random uncertainty estimates from Section~{\ref{sec:error_fwd_model}}}}\label{fig:f107_ap_texo_scatter}
\end{figure}

SUVI occultation density and temperature measurements are currently available from the September equinox 2018 occultation season, starting 2018-08-31, through the present. These data therefore span solar minimum between Solar Cycle 24 and Solar Cycle 25 along with the present rise of Solar Cycle 25. The GOES-R series satellites are projected to operate through into the mid-2030s, thus \change{likely}{potentially} providing a continuous climatological record of this type for the entirety of Solar Cycle 25. \change{ Figures {\ref{fig:climate_composition}} and {\ref{fig:climate_temp}} show}{Figure {\ref{fig:climate_composition}} shows} the compositional density measurements\remove{ and temperature measurements} from GOES-16/SUVI from the September equinox 2018 through \change{March}{September} equinox 2024 occultation seasons, along with the F$_{10.7}$ radio flux representing the solar activity level. As previously mentioned, and as show in Figure \ref{fig:occ_lats}, each occultation season covers latitudes from pole to pole for both the dawn and dusk measurements. \add{This latitudinal sampling for each occultation season causes the scalloping pattern seen in the long-term O and N$_2$ densities of Figure {\ref{fig:climate_composition}} which are otherwise not driven by solar activity or geomagnetic activity.}\add{ Figure {\ref{fig:climate_temp}} shows the time series of exospheric temperature measurement for all currently available occultation seasons along with the F$_{10.7}$ cm radio flux solar forcing (Figure {\ref{fig:climate_temp}}a) and geomagnetic Ap index (Figure {\ref{fig:climate_temp}}b).} 
\add{ Also, shown in Figure {\ref{fig:f107_ap_texo_scatter}} are scatter plots of SUVI derived exospheric temperature with respect to  F$_{10.7}$ flux (Figure {\ref{fig:f107_ap_texo_scatter}}a \& b) and Ap (Figure {\ref{fig:f107_ap_texo_scatter}}c \& d) for observations at high latitudes ($\left|\mathrm{Lat}\right|>60$) and middle to low latitudes ($\left|\mathrm{Lat}\right|\leq60$), along with the Pearson correlation (r-value in plot) between the forcing index and temperature for each group. As expected, there is a strong corelation for all observations (r-value $\approx$ 0.9) with exospheric temperature and F$_{10.7}$ as solar EUV is the primary heating source of the thermosphere. Figure {\ref{fig:f107_ap_texo_scatter}}c \& d show moderate correlation between exospheric temperature and Ap (r-value $\approx$ 0.3). Notably, for dusk observations, a two-tail z-test did not reveal a statistically significant difference between the high latitude measurements and the mid-to-low latitude measurements, at the 5\% confidence level, when controlling for solar activity (p-value $=0.57>0.05$ and p-value $=0.23>0.05$ for high solar activity and low solar activity, respectively). However, at dawn there is a statistically significant difference between high latitude and mid-to-low latitude measurements (p-value $=\num{2.6e-12}<0.05$ and p-value $=\num{6.5e-14}<0.05$ for high solar activity and low solar activity, respectively), and as such have separate regressions and correlation coefficients, shown in purple.}

\subsection{Storm-time Response}\label{sec:storm}
\begin{figure}[H]
    \centering
    \includegraphics[width=.95\linewidth]{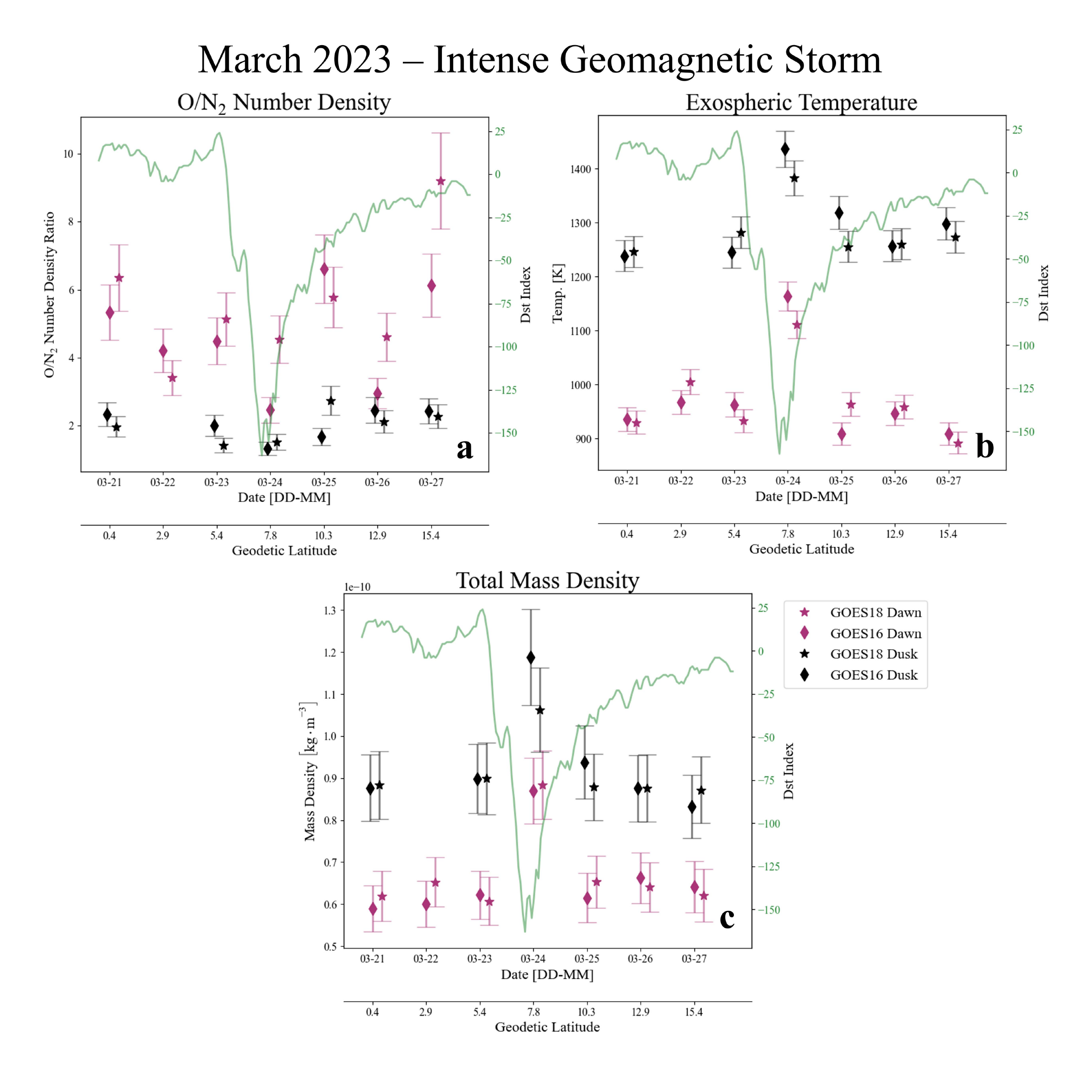} 
    \caption{\remove{\textbf{(a)} GOES-16 \& GOES-18/SUVI occultation derived total mass density enhancement at 250 km at the dusk terminator during the 24 March 2023 intense geomagnetic storm (blue) and corresponding Dst index (orange). \textbf{(b)} Same as \textbf{(a)} but for the dawn terminator observations. Measurements were observed at latitudes from $\sim$0 -- 15\textdegree N with observations at the peak main phase occurring around 7.5\textdegree N. Error bars are calculated from the 1-$\sigma$ LOS density errors, as discussed in Section {\ref{sec:error_fwd_model}.}}\add{{\textbf{(a)}} GOES-16 \& GOES-18/SUVI occultation derived O/N$_2$ ratio at 250 km during the 24 March 2023 intense geomagnetic storm at dusk (black) and dawn (purple) terminators with the corresponding Dst index (green). \textbf{(b)} Same as \textbf{(a)} but for for exospheric temperature. \textbf{(c)} Same as in \textbf{(a)} but for total mass density. All error bars represent the random uncertainty estimates from Section~{\ref{sec:error_fwd_model}}}.}\label{fig:storm}
\end{figure}

SUVI occultation measurements provide dusk and dawn terminator measurements roughly one hour apart, meaning for a geomagnetic storm-time event both the dawn and dusk thermospheric response at the same latitude can be observed in near-succession. Furthermore, with measurements from two GOES-R satellites in each the GOES-East and GOES-West position, as with GOES-16 and GOES-18, measurements of the same latitude and occultation terminator, can be observed roughly 4 hours apart for thermospheric storm-time studies and prediction. Figure \ref{fig:storm} shows the \add{O/N$_2$ ratio, exospheric temperature and }total mass density \add{at 250 km} derived from GOES-16 and GOES-18 SUVI occultations during the 24 March 2023 geomagnetic storm. During this event, SUVI occultations show a dusk side terminator mass density enhancement of 32\% and dawn side enhancement of 46\% at the peak of the storm's main phase over the pre-storm values. The disturbance storm time (Dst) index shows a main phase storm peak around 03:00 UT and SUVI dusk side measurements peak at the GOES-16 measurement around 04:30 UT and begin to recover at the GOES-18 measurement around 08:30 UT. However, the dawn measurements show a peak enhancement at the 09:50 UT GOES-18 measurement, after the dusk side orbit recovery has started. Both the dawn and dusk terminators' \change{mass densities}{states} recover to their pre-storm values by the 25 March observations, while the Dst shows the storm is still in the recovery phase.

\section{Comparisons}\label{sec:Comps}
Any comparisons between SUVI measured neutral densities and other direct observations are scarce, due primarily to the dearth of observations of this type at these altitudes in the thermosphere, but also any of the few measurements of this type that exist must also coincide temporally and spatially with the SUVI observations. Due to these difficulties, we instead present in this section comparisons with tools commonly used for space weather forecasting and now casting: empirical and assimilative models.
\subsection{NRLMSIS-2.0}
\label{sec:msis_comp}
\begin{figure}[H]
    \centering
    \includegraphics[width=.95\linewidth]{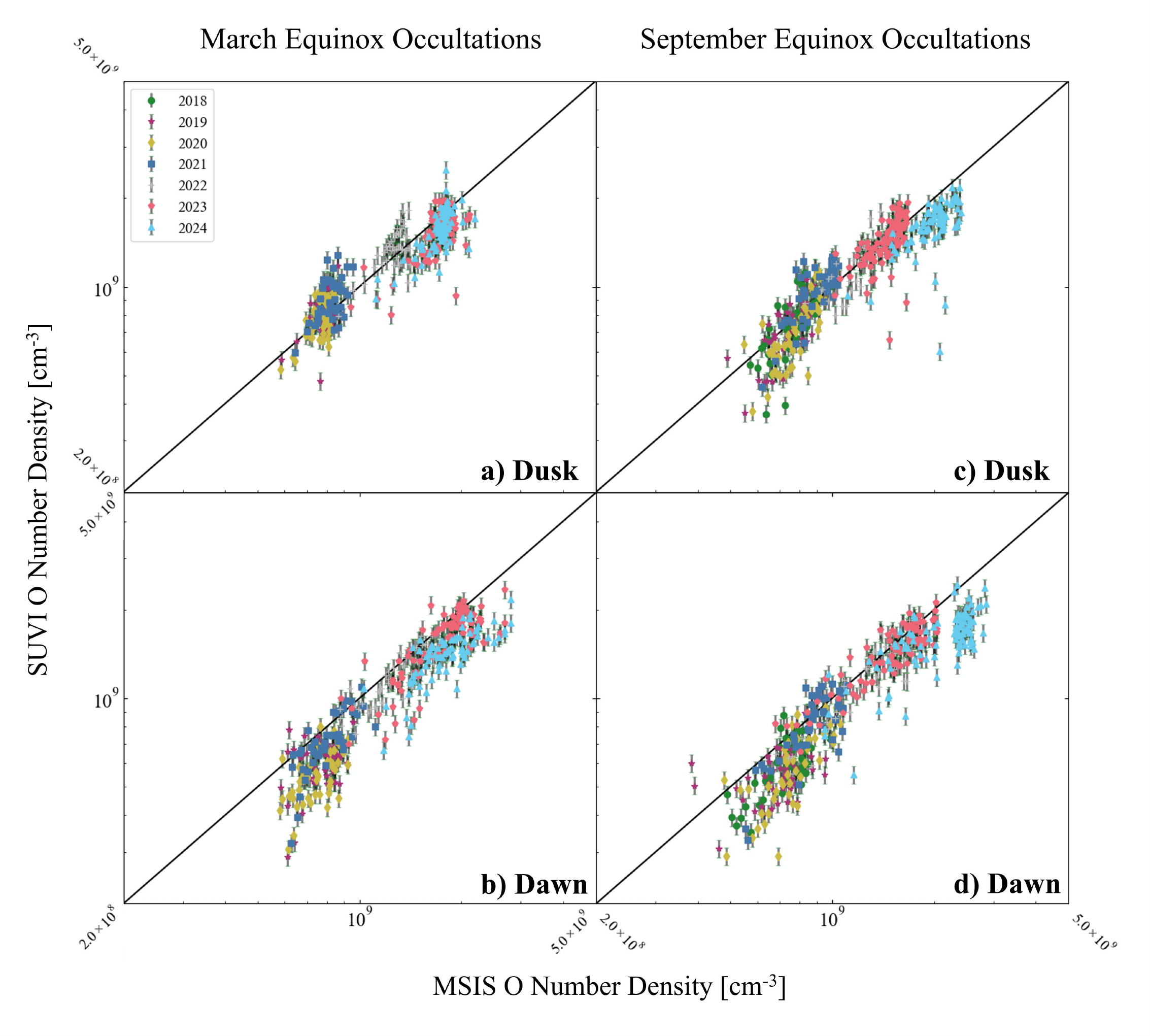} 
    \caption{\textbf{(a)} SUVI retrieved \change{total (O+N$_2$)}{O} \change{LOS}{number} densities \change{at 250 km}{with 250 km tangent height} for all \change{MArch}{March} equinox occultation seasons at dusk local time vs. MSIS reported LOS densities at the same observing time and location. \textbf{(b)} same as in \textbf{(a)} but for all dawn local time observations. \textbf{(c)} Same as in \textbf{(a)} but for all September equinox occultation seasons. \textbf{(d)} Same as in \textbf{(a)} but for all September equinox occultation seasons at dawn local time. The black curve in all four plots depicts the conceptual 1-to-1 agreement line between data sets. \add{All error bars represent the random uncertainty estimates from Section~{\ref{sec:error_fwd_model}}}}\label{fig:MSIS_comp_o}
\end{figure}

\begin{figure}[H]
    \centering
    \includegraphics[width=.95\linewidth]{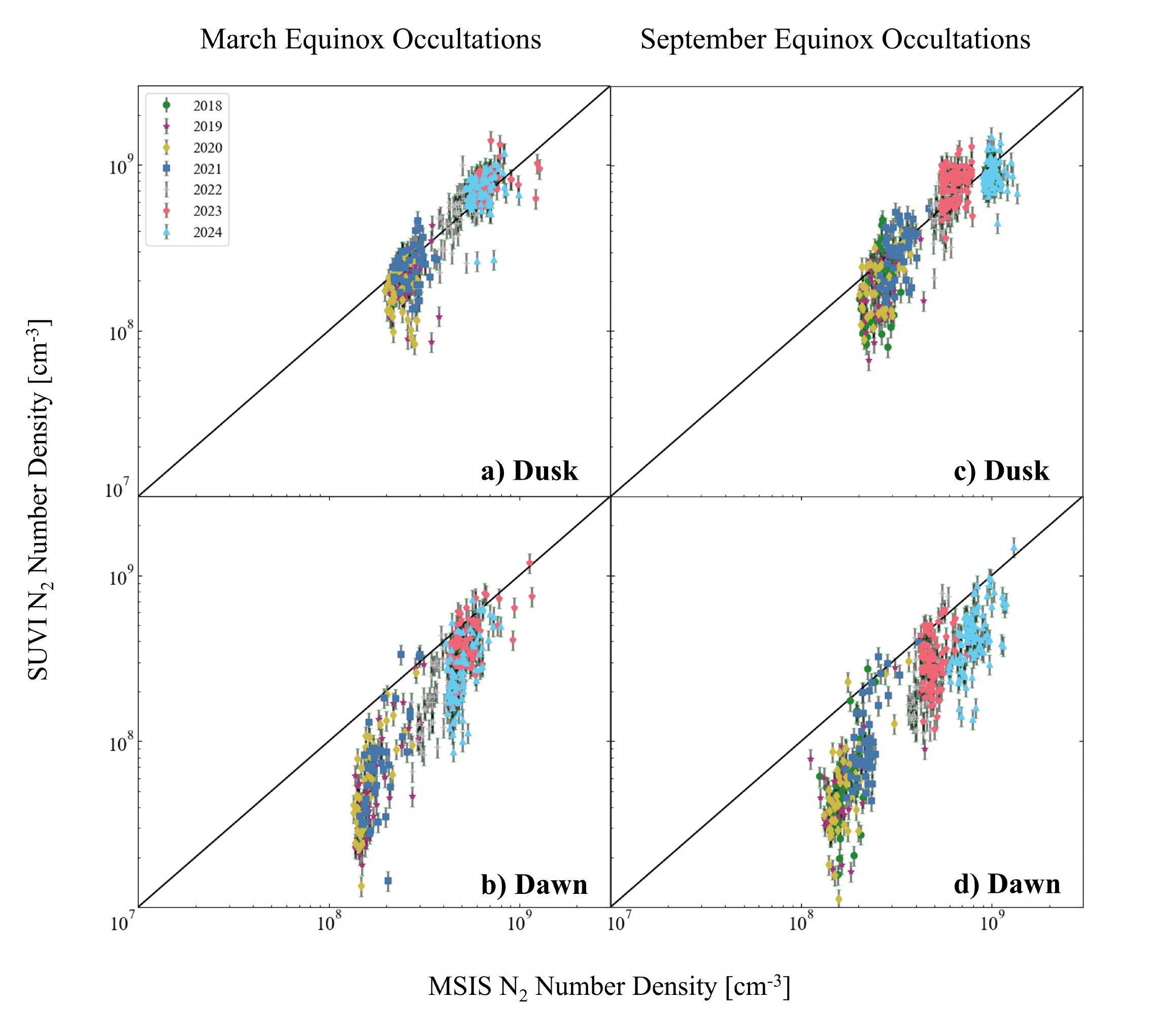} \caption{\textbf{(a-d)} Same as in Figure \ref{fig:MSIS_comp_o} \textbf{(a-d)}, respectively, but for N$_2$ number densities derived from SUVI occultations compared to MSIS.}\label{fig:MSIS_comp_n2}
\end{figure}

\begin{figure}[H]
    \centering
    \includegraphics[width=.95\linewidth]{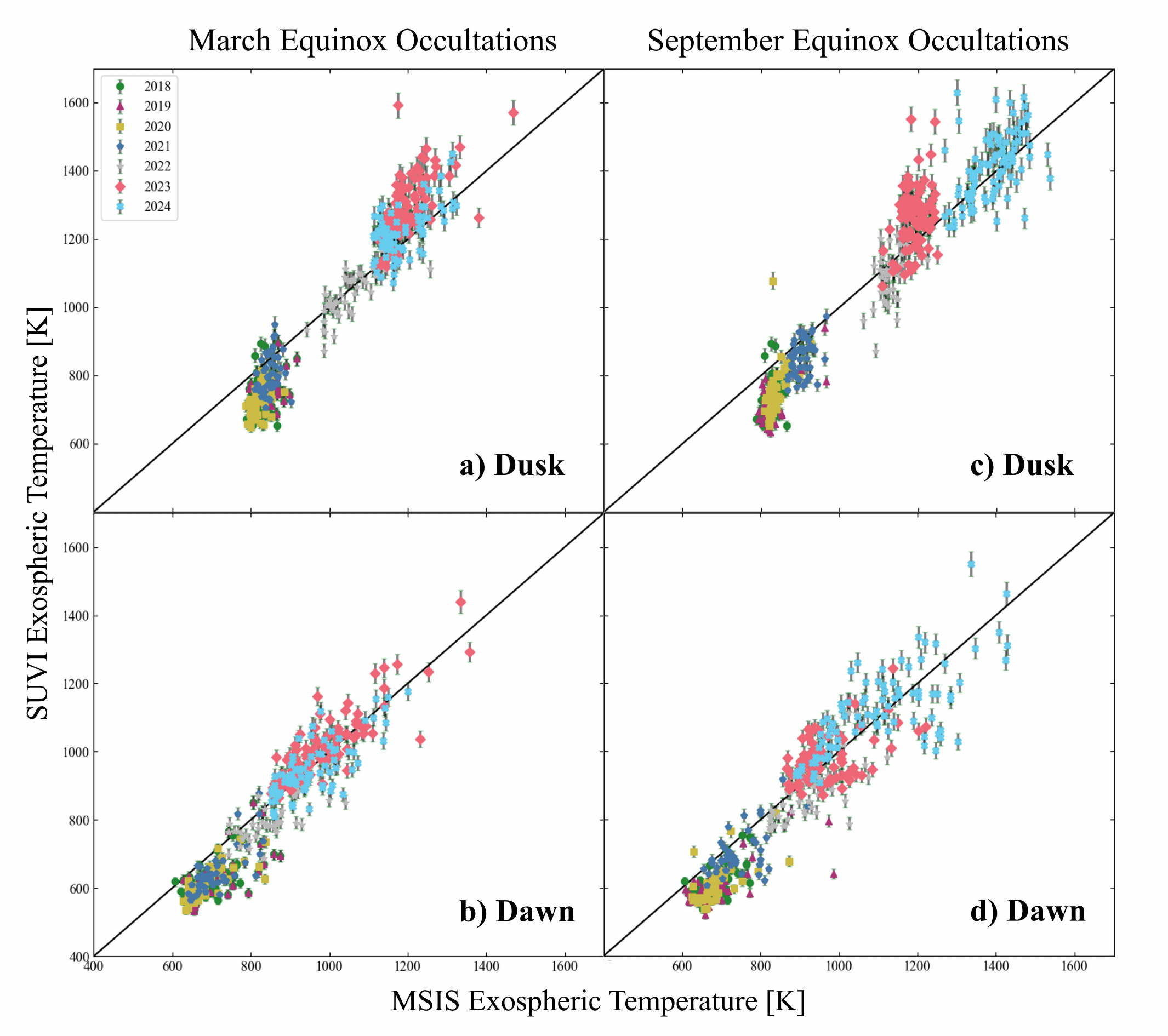} 
    \caption{\textbf{(a-d)} Same as in Figure \ref{fig:MSIS_comp_o} \textbf{(a-d)}, respectively, but for exospheric temperature derived from SUVI occultations compared to MSIS.}\label{fig:MSIS_comp_temp}
\end{figure}

NRLMSIS-2.0 model is an empirical atmospheric model which provides temperature and neutral densities of the dominant atmospheric species from the ground to the exobase \citep{MSIS}. NRLMSIS-2.0, and \change{it's}{its} predecessor NRLMSISE-00 \citep{MSIS00}, are commonly used for space weather applications, and are driven by F$_{10.7}$ radio flux and geomagnetic Ap index. Figure {\ref{fig:MSIS_comp_o}} shows all SUVI occultation O \change{LOS}{number} density measurements and Figure {\ref{fig:MSIS_comp_n2}} shows all SUVI occultation N$_2$ number density measurements, at a tangent altitude of 250 km, compared to MSIS reported number densities during the same SUVI observation times and geometries. Also, Figure \ref{fig:MSIS_comp_temp} shows exospheric temperature measurements from SUVI occultations compared to MSIS. \change{While, in general, there is reasonable agreement between SUVI number density and temperature measurements to the MSIS model outputs, there appear to be two noticeable discrepancies: 1) SUVI dawn measurements are generally less than the MSIS reported values, 2) for seasons during solar minimum (seasons in 2018-2021) SUVI measurements are less than the MSIS reported values with the relationship between SUVI and MSIS density estimate in Figure {\ref{fig:MSIS_comp_o}} \& {\ref{fig:MSIS_comp_n2}} deviating from a linear relationship.}{Overall, SUVI number density and temperature measurements show reasonable agreement with MSIS model outputs. However, two notable discrepancies are observed: (1) SUVI dawn measurements are generally lower than MSIS values, and (2) during solar minimum seasons (2018–2021), SUVI measurements tend to underestimate MSIS densities, with the relationship shown in Figures {\ref{fig:MSIS_comp_o}} \& {\ref{fig:MSIS_comp_n2}} deviating from linearity.} Table \ref{table:msis_percent_diff} shows the percent difference between SUVI and MSIS O and N$_2$ number densities and total mass density at 250 km, as well as differences in exospheric temperature, each broken into occultation season and solar activity. This shows that for all quantities\remove{ -- except for Dusk March equinox occultation observations --} the differences between the MSIS model and SUVI measurements is larger for low solar activity and that all dawn SUVI measurements are lower than what MSIS reports.\remove{ These discrepancies are discussed further in Section {\ref{sec:summary}}.}
    

\begin{table}[H]
\begin{center}
\caption{Percent difference between SUVI and MSIS O and N$_2$ \change{LOS}{number} densities at an altitude of 250 km, total mass density \change{at 250 km}{at an altitude of 250 km}, and exospheric temperature for all seasons (total) for seasons during solar minimum (F$_{10.7}$$<$100 as a proxy) and non-solar minimum (F$_{10.7}$$\geq$100 as a proxy).}\label{table:msis_percent_diff}
\begin{tabular}{c c c c c c c c c }
\multicolumn{9}{c}{SUVI Occulations to MSIS Model}\\
\multicolumn{9}{c}{Percent Differences}\\
\toprule
     \multicolumn{9}{c}{O Number Density}\\
   \cmidrule(lr){2-8}
   &&\multicolumn{3}{c}{March Equinox Seasons} & \multicolumn{3}{c}{September Equinox Seasons}&\\
   && Total& F$_{10.7}$$<$100&F$_{10.7}$ $\geq$100& Total&F$_{10.7}$$<$100& F$_{10.7}$$\geq$100&\\
    &Dusk& 0.7 & 7.2 & -3.9 & -1.6 & 1.0 & -3.7 &\\
      &Dawn & -14.5 & -17.1 & -12.6 & -11.8 & -14.5 & -9.8&\\
       \multicolumn{9}{c}{N$_2$ Number Density}\\
       \cmidrule(lr){2-8}
       &&\multicolumn{3}{c}{March Equinox Seasons} & \multicolumn{3}{c}{September Equinox Seasons}&\\
       && Total& F$_{10.7}$$<$100&F$_{10.7}$ $\geq$100& Total&F$_{10.7}$$<$100& F$_{10.7}$$\geq$100&\\
       &Dusk& -3.0 & -22.2 & 10.7 & -9.7 & -23.0 & 1.0&\\
      &Dawn & -51.5 & -65.7 & -41.4 & -54.9 & -66.6 & -46.0&\\
      \multicolumn{9}{c}{Total Mass Density}\\
       \cmidrule(lr){2-8}
       &&\multicolumn{3}{c}{March Equinox Seasons} & \multicolumn{3}{c}{September Equinox Seasons}&\\
       && Total& F$_{10.7}$$<$100&F$_{10.7}$ $\geq$100& Total&F$_{10.7}$$<$100& F$_{10.7}$$\geq$100&\\
       &Dusk& 1.1 & -2.3 & 3.3 & -3.0 & -6.6 & 0.0&\\
      &Dawn & -24.2 & -29.7 & -20.3 & -24.8 & -28.5 & -22.1&\\
      \multicolumn{9}{c}{Exospheric Temperature}\\
       \cmidrule(lr){2-8}
      &&\multicolumn{3}{c}{March Equinox Seasons} & \multicolumn{3}{c}{September Equinox Seasons}&\\
   && Total& F$_{10.7}$$<$100&F$_{10.7}$ $\geq$100& Total&F$_{10.7}$$<$100& F$_{10.7}$$\geq$100&\\ 
    &Dusk& -2.8 & -10.4 & 4.2 & -3.9 & -11.3 & 1.8 &\\
      &Dawn & -5.7 & -11.9 & -0.1 & -4.8 & -10.9 & -0.1&\\
\bottomrule
\end{tabular}
\end{center}
\end{table}

\subsection{Assimilative Model Comparisons}\label{sec:assim}
The Iterative Driver Estimation and Assimilation (IDEA) is an assimilative model that adjusts the energy inputs into the Thermosphere-Ionosphere-Electrodynamics General Circulation Model (TIE-GCM) based on in-situ mass density data, such as that from the Challenging Mini-Satellite Payload (CHAMP) and Gravity Recovery And Climate Experiment (GRACE) satellites \citep{IDEA}. The IDEA-GRACE-FO is an adaption to this model that also assimilates the GRACE Follow-On (GRACE-FO) accelerometer data inferred mass densities. 

\begin{figure}[H]
    \centering
    \includegraphics[width=\linewidth]{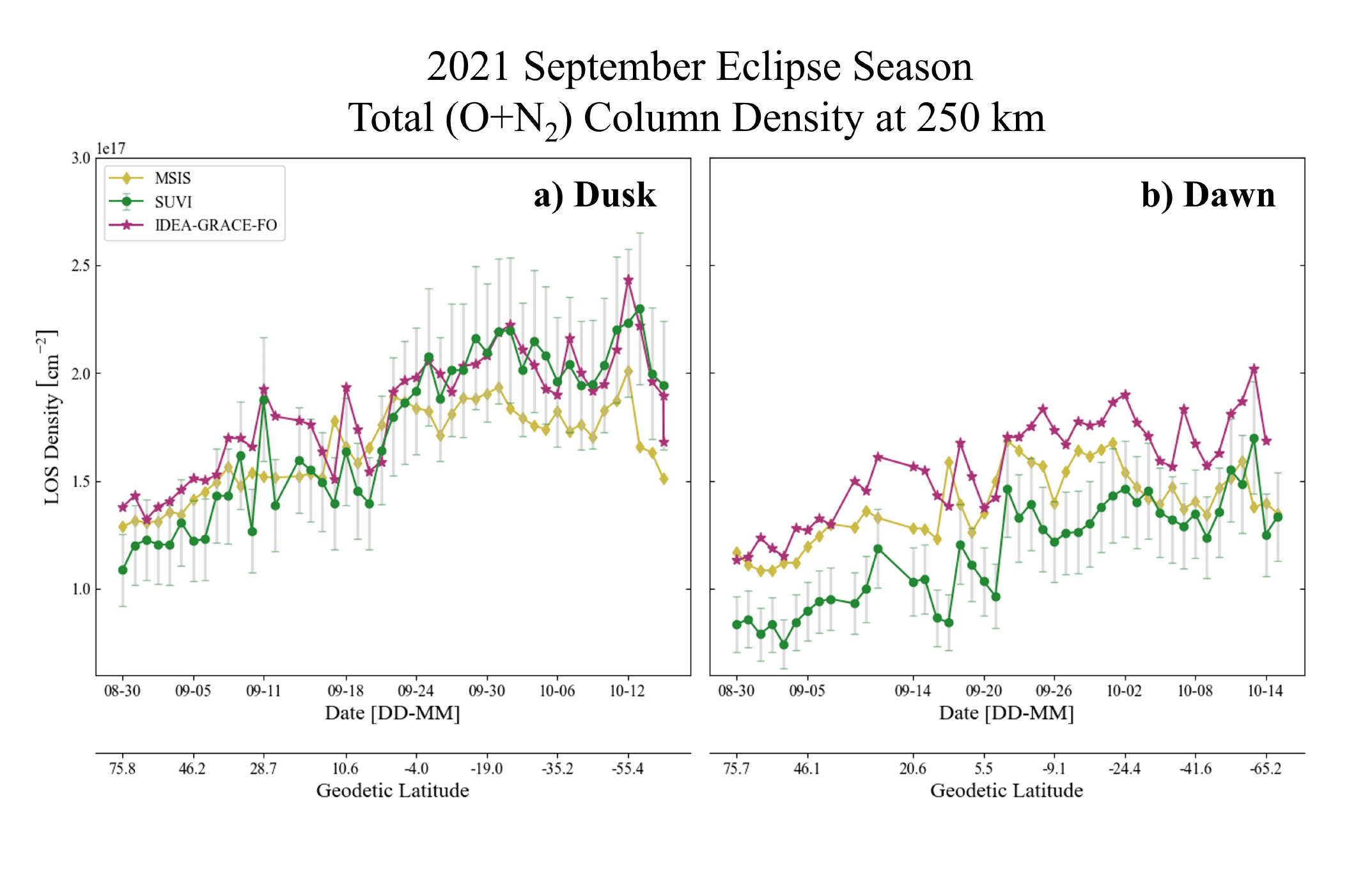} 
    \caption{\textbf{(a)} 2021 September equinox SUVI dusk occultation season total (O+N$_2$) LOS density measurements \change{at 250 km}{with a tangent ray height of 250 km} (\change{orange}{green}-dot) compared to the IDEA-GRACE-FO assimilative model (\change{blue}{purple}-star) and MSIS model (\change{green}{yellow}-diamond). \textbf{(b)} Same as in \textbf{(a)} but for dawn observations. \add{All error bars represent the random uncertainty estimates from Section~{\ref{sec:error_fwd_model}}}}\label{fig:IDEA_comp}
\end{figure}

\begin{figure}[H]
    \centering
    \includegraphics[width=\linewidth]{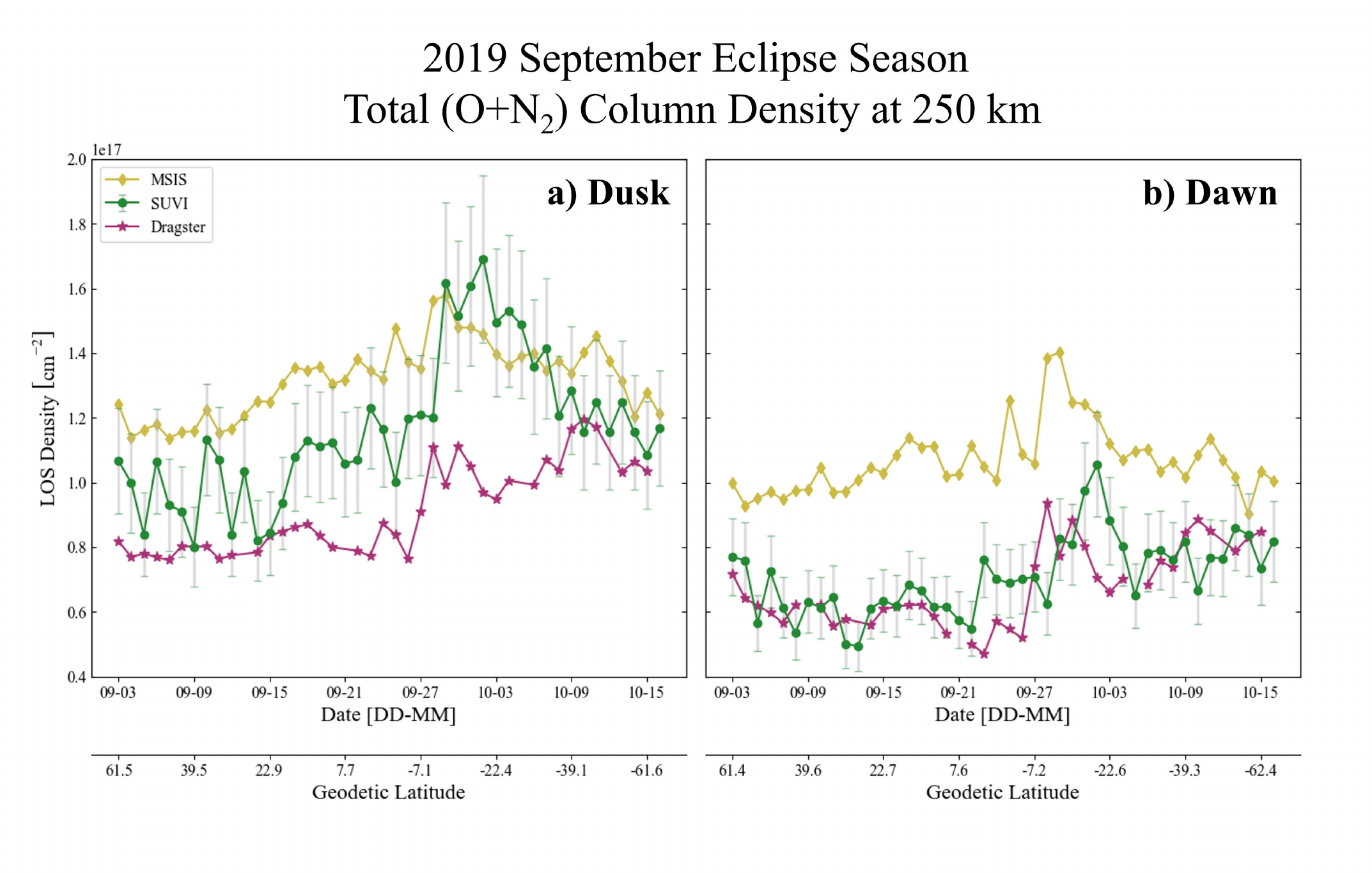} 
    \caption{\textbf{(a)} 2019 September equinox SUVI dusk occultation season total (O+N$_2$) LOS density measurements (\change{orange}{green}-dot) compared to the Dragster assimilative model (\change{blue}{purple}-star) and MSIS model (\change{green}{yellow}-diamond) at a 250 km tangent ray altitude. \textbf{(b)} Same as in \textbf{(a)} but for dawn observations. \add{All error bars represent the random uncertainty estimates from Section~{\ref{sec:error_fwd_model}}}}\label{fig:Dragster_comp}
\end{figure}

Shown in Figure \ref{fig:IDEA_comp} are comparisons of SUVI measured total LOS density \change{at 250 km}{with a tangent ray height of 250 km} to the MSIS\remove{-2.0} empirical model and the IDEA-GRACE-FO assimilative model outputs during the 2021 September equinox occultation season, for which there is temporal overlap of the GRACE-FO assimilated data and SUVI measurements. During this temporal overlap, the GRACE-FO was operating at $\sim$500 km with sampling mostly aligned with dawn/dusk. At the dusk terminator IDEA-GRACE-FO outputs and SUVI measurements are in good agreement, with an average percent difference \change{-2.02}{-2}\%, however -- as seen with the MSIS comparisons in Section \ref{sec:msis_comp} -- the dawn terminator shows a much larger disagreement, with an average percent difference of \change{-23.67}{-24}\%.
Dragster is an assimilative model that assimilates orbital data 
from 70-100 LEO space objects into an ensemble of general circulation or empirical models \citep{Pilinski2019, Pilinski2016}. Figure \ref{fig:Dragster_comp} shows comparisons of SUVI measured total LOS density \change{at 250 km}{for a 250 km tangent ray height} to the MSIS\remove{-2.0} empirical model and the Dragster assimilative model outputs during the 2019 September equinox occultation season. For the time period shown here, assimilated data spans from approximately 200 km to 700 km altitude. The assimilation used 60 NRLMSISE-00 \change{ensembel}{ensemble} members. The estimated state included 6-hour geomagentic (Ap) and 1-day solar forcing (F107) alongside gridded corrections to the output model density at two different altitudes. Unlike the comparisons between SUVI and MSIS or SUVI and IDEA-GRACE-FO, the Dragster model shows better agreement with SUVI at dawn, with a percent difference of \change{+1.76}{+2}\%, than at dusk, with a percent difference of \change{+12.88}{+13}\%. Furthermore, for the dusk measurements at 250 km\add{ tangent ray} altitude, the agreement between the Dragster pre-assimilation model LOS densities and SUVI was 38\% linear standard deviation, which was reduced to 20\% with assimilation.
Similarly, for the dawn measurements at 250 km \add{tangent ray} altitude, the agreement between the pre-assimilation model LOS densities and SUVI was 50\% linear standard deviation, which was reduced to 23\% with assimilation.
For both dawn and dusk, the SUVI-Dragster model standard deviation was reduced by approximately a factor of two when drag-derived density data is assimilated into Dragster. This improvement in SUVI-model agreement indicates that the time-series information in the SUVI dataset is broadly consistent with the orbit-averaged densities being assimilated by Dragster.

Both the IDEA-GRACE-FO and Dragster assimilative models use inferred in-track mass density from accelerometer data or energy dissipation rates estimated from ground-based observations, respectively. While these measurements may provide greater spatial coverage than SUVI occultations can provide, they are fixed only at the operating altitude of the satellites. Due to this fact, orbital drag and accelerometer data is historically rare and practically short lived at altitudes below 300 km, making occultation products a valuable way to persistently provide altitude resolved information for assimilative schemes.  Specifically, total LOS O and N$_2$ number densities can be assimilated into these models, which are comparable to the along-track drag measurements that these models typically assimilate.

\section{Discussion}\label{sec:Discussion}

SUVI solar occultation observations are being used to produce O, N$_2$, and temperature profiles from $\sim$180-500 km. As discussed in Section \ref{sec:error_fwd_model}, the retrieval forward model method used to derive these profiles, from the fundamental ER measurement, are reported here to have random uncertainty\change{ at 250 km}{, at a tangent height of 250 km,} of \change{6.86}{8}\%, \change{13.68}{17}\% and \change{2.30}{3}\% for the O \change{LOS}{number} density, N$_2$ \change{LOS}{number} density and temperature, respectively. As discussed in Section \ref{subsec:CS_Uncertainty}, minor adjustments to the total effective absorption cross sections 
over the SUVI 17.1 nm and 30.4 nm passbands were required to improve the retrieval's performance. These adjustments were to the relative offset between the 17.1 nm and 30.4 nm to the 19.5 nm total effective cross section, while keeping the absolute value of 19.5 nm effective cross section fixed. \change{These adjustments are at most $\sim$5\% and within the lab cross section uncertainty, however, these small changes greatly improved the density retrieval self-consistency, by reducing the standard deviation between comparative measurements by  $\sim$25\% for O and $\sim$45\% for N$_2$}{These adjustments are at most $\sim$5\% and remain within the laboratory cross section uncertainty; however, even these small changes significantly improved the self-consistency of the density retrievals, reducing the standard deviation between comparative measurements by $\sim$25\% for O and $\sim$45\% for N$_2$ compared to using the unadjusted lab cross sections}. \add{Recent theoretical cross sections by \mbox{\citet{Soto2024}} were also considered in the analysis as an alternative to the \mbox{\citet{Fennelly1992}} cross sections. However, these cross sections resulted in poorer direct retrieval LOS densities than those derived from the lab cross sections from \mbox{\cite{Fennelly1992}} and the adjusted cross sections with $\sim$15\% more non-physical retrievals (e.g. negative densities profiles of a species) and, as such, were not used in the forward model in place of the adjusted cross sections. Furthermore, these theoretical cross sections do not provide associated uncertainties preventing a meaningful assessment of their differences when compared to the lab or adjusted cross sections used in this study.} \add{Section~{\ref{sec:cs_error}} demonstrates the impact uncertainties in the effective cross section have on solar occultation measurements of this type, that for even small cross section uncertainties ($\sim$5\% in each channel) can manifest as systematic biases on the order of 50\% for O and 100\% for N$_2$ retrieved densities. However, despite these potentially large compositional biases, the retrieved mass densities and temperature are largely unaffected with potential systematic biases on the order of 10\% and 20\% respectively and are relatively uniform with altitude (see Figure~{\ref{fig:cs_offsets}}d \& e). Furthermore, the O/N$_2$ composition ratio bias scenarios (Figure~{\ref{fig:cs_offsets}}c) also exhibit relatively uniform offsets with altitude. This highlights that although uncertainties in the effective cross sections can introduce systematic biases, the absolute mass density remains largely unaffected for satellite drag applications, and relative, temporal variations in the O/N$_2$ ratio remain reliable for ionospheric space weather studies.}\change{ Other}{ However, other} studies have discussed the need for more accurate measurements of the photoabsorption cross sections of thermospheric major species \citep[see][]{Lumpe2007}, and future occultation measurements would greatly benefit from such data\add{ for retrieving absolute, computationally resolved density profiles.}\remove{ especially at the temperatures seen in the thermosphere}. 

As discussed in Section \ref{sec:climate}, this new SUVI solar occultation data set spans from 2018 through the present, and is expected to be operational through the GOES-R series mission through the mid-2030s. This data will provide a climatological record of the thermospheric state through Solar Cycle 25. \add{Figure {\ref{fig:f107_ap_texo_scatter}} showed exospheric temperature as a function of forcing from solar irradiance, represented by F$_{10.7}$ flux, and geomagnetic activity, represented by Ap. These relationships, show stronger correlation between exospheric temperature and solar activity than that with geomagnetic activity, as expected. At high latitudes the distribution of dusk temperatures is not statistically different from those at mid-to-low latitudes, however, a statistical difference is seen at dawn . This potentially indicates that heating at the poles at dusk is dominated mostly by EUV heating, while at dawn other heating sources at the poles, such as that from particle precipitation and Joule heating, play a more significant role in driving thermospheric temperature. In addition, the relationship between exospheric temperature and F$_{10.7}$ flux, shows a leveling off in the temperature response at high levels of solar activity, deviating from a linear relationship. This has been seen previously with MSIS and the measurements it is based on \mbox{\citep[see][]{Hedin1987}} as well as with the TIMED/GUVI exospheric temperature measurements \mbox{\citep[see][]{Zhang2011}} and seems to suggest that the dependence of thermospheric temperature on solar EUV irradiance decreases at high irradiance levels.} \change{Furthermore, as shown in Section {\ref{sec:storm}}, as SUVI occultations are being taken on the multiple GOES-R satellites, successive measurements by each satellite allows for studies of smaller timescale thermospheric dynamics such as those during geomagnetic storms.}{In addition to climatological data, SUVI occultations are useful to studying smaller timescale thermospheric dynamics. As measurements are taken from both GOES-East and GOES-West in near succession at the same latitude and local solar time, storm-time dynamics can be assessed, as discussed in {\ref{sec:storm}}, providing useful diagnostics for future space weather operations.}

From Section \ref{sec:Comps}, SUVI retrieved densities show reasonable agreement\add{, within $\sim$30\% over all compared measurements}, with MSIS empirical model and the IDEA-GRACE-FO and Dragster assimilative models. Noticeable differences are seen between MSIS and IDEA-GRACE-FO when compared to SUVI retrieved total LOS densities at dusk. For much of the data these differences are outside the error estimates for retrieved total LOS density as reported in Section \ref{sec:error_fwd_model}. The 2019 September equinox seasons comparisons between SUVI and Dragster, however, do not show this discrepancy in the dawn measurements, but instead show a similar magnitude discrepancy at dusk. While a given GOES-R satellite is experiencing strong thermal gradients upon eclipse exit during the dawn measurements, this dawn discrepancy is unlikely due to instrumental effects on SUVI as the pole-ward observations are taken nearly at the same universal time as the satellite scans the pole terminator, meaning there is effectively little to no fully eclipsed part of the orbit, yet these poleward observations still exhibit this discrepancy. If this was due to thermal gradients of the instrument there would be a latitudinal dependence of the discrepancy, as orbits in the middle of the occultation season (measurements near Earth's equator), would have longer eclipses and steeper thermal gradients, but this latitudinal dependence is not seen in the data.  Moreover, this discrepancy is seen in both GOES-16 and GOES-18 SUVI occultation retrievals with a similar magnitude, thus also making individual instrument effects unlikely to be the cause. \change{Furthermore, this is unlikely an undiscovered systematic error in forward modeling the atmosphere, as the dawn and dusk measurements are treated in the same manner, yet the dusk observations do not show this same magnitude of offset from the modeling results.}{If this discrepancy is due to an undiscovered systematic error in the forward modeling process, this error must be biased to dawn-side occultations, despite both terminators are handled the same way in processing.} \add{For this reason, it is also unlikely that the discrepancy arises from error in the adjusted total effective cross section used in this study, as such an effect would be systematic and would impact both dawn and dusk measurements similarly.} Future work to better understand this discrepancy will be to study more occultation seasons compared to the assimilative models in Section \ref{sec:assim}. \change{In particular, the better agreement at dawn between Dragster and SUVI may be due to the fact that the data assimilated covers altitude ranges closer to the range of the SUVI occultation observations. The dawn-dusk discrepancy in the SUVI-Dragster comparison, may be due to variations in assimilated satellite local-time and latitude coverage. Data assimilated by Dragster below 300 km altitude comes from a few satellites in elliptical orbits. As such, the local time and latitude range at which these orbits most interact with the thermosphere changes on timescales of weeks and months, leading to variability in sampling. In addition, the GRACE-FO accelerometer data that was assimilated for the IDEA-GRACE-FO to SUVI comparisons was taken around 490 km, which lies above the majority of the SUVI observing altitudes. If it is found that the lower altitude data that went into the Dragster assimilation was sampled more near dawn and is in better agreement with SUVI measured densities, it would be compelling evidence that there is truly a lower average density at the dawn terminator than what both MSIS and TIE-GCM (the base model data for the IDEA assimilative model) predict.}{In particular, we speculate that the better agreement at dawn between Dragster and SUVI during the September 2019 occultation season is due to the fact that the data assimilated covers altitude ranges closer to the range of the SUVI occultation observations and a greater number of drag measurements at dawn. Data assimilated by Dragster below 300 km altitude comes from a few satellites in elliptical orbits. As such, the local time and latitude range at which these orbits most interact with the thermosphere changes on timescales of weeks and months, leading to variability in latitude and local time sampling. In addition, the GRACE-FO accelerometer data that was assimilated for the IDEA-GRACE-FO to SUVI comparisons was taken from a circular orbit around 490 km, which lies above the majority of the SUVI observing altitudes. Because the lower altitude data that went into the Dragster assimilation was sampled more near dawn and is in better agreement with SUVI measured densities, this may imply a lower average LOS density at the dawn terminator than what both MSIS and TIE-GCM (the base model data for the IDEA assimilative model) predict. However, future comparisons should be between number densities derived by each product rather than LOS density (to rule out bias by atmospheric species) and should be made across more occultation seasons.}

Furthermore, from Section \ref{sec:msis_comp} and Figures \ref{fig:MSIS_comp_o} \& \ref{fig:MSIS_comp_n2}, MSIS and SUVI densities show worsened agreement during quiet solar conditions. The original mass spectrometer composition and density data that went into the first MSIS series model (NRLMSIS-77) were taken during times with the F$_{10.7}$ $ \gtrsim 80$, therefore only data taken when $100 >$ F$_{10.7}$ $\gtrsim 80$ is used for the lowest data sorting box F$_{10.7}$ $<100$ \citep[see][]{MSIS77}; meaning MSIS-77 may be biased high during times when F$_{10.7}<80$.  While more datasets have gone into subsequent MSIS model updates, these earlier satellite mass spectrometer data and later rocket mass spectrometer data (which comprise most of the N$_2$ and O density inputs to MSIS) do not include measurements for F$_{10.7}<80$ \citep[see][]{MSIS77,MSIS86}. This potential under-sampling of quiet solar times in MSIS input data, in addition to the particularly quiet solar minimum between Solar Cycle 24 and 25 observed during SUVI occultation seasons in 2018-2021, may cause the discrepancies seen for times of low solar activity. \add{Moreover, this discrepancy may be attributed to the secular decrease in density observed by \mbox{\citet{Emmert2004}} which would cause a high bias in the MSIS reported densities at solar minimum \mbox{\citep{Emmert2008}} for this solar minimum observed by SUVI.} \change{Moreover}{Lastly}, MSIS and TIE-GCM use F$_{10.7}$ as a proxy for EUV irradiance. However, at solar minimum F$_{10.7}$ becomes non-linear with EUV irradiance \citep{Solomon2011}, perhaps causing models that are driven by this proxy to overestimate the thermospheric density and temperature. If these observed discrepancies are caused by real differences, densities from SUVI occultations provide crucial measurements of the thermospheric state during quiet solar conditions\remove{ not otherwise observed}, which may be invaluable for validation of and assimilation into empirical and physics-based models.
\section{Summary}\label{sec:summary}
SUVI solar occultations provide number density measurements of the dominant thermospheric species O and N$_2$, as well as neutral temperature, from September of 2018 to the present. The technique used to derive the thermospheric state from solar EUV occultation images, to the authors' knowledge, is novel to this application -- but the same methods could be applied to the increasing number of solar EUV imagers in Earth orbit. 

Further investigation is required to study discrepancies between observed absorption cross sections and lab measurements, as discussed in Section \ref{subsec:CS_Uncertainty}. Also requiring further investigation are the discrepancies seen between dawn terminator measurements by SUVI and model predictions by the Dragster assimilative scheme and those reported from the MSIS empirical model and IDEA-GRACE-FO assimilative method, discussed in Section \ref{sec:Comps} and \ref{sec:Discussion}. 

The density profiles of O and N$_2$ and neutral temperature profiles are being produced as part of the nominal SUVI L2 data pipeline. This data will be available through the GOES-R series mission life, providing a much-needed climatological record of the ``Thermospheric Gap" in low earth orbit required for space weather model validation and data assimilation. No measurements of the thermospheric state are currently being made in real-time at the observing altitudes of SUVI occultations, however, as these measurements only require the already produced real-time NOAA space weather L1b SUVI images, these density and temperature profiles could be produced
in real-time (rather than the nominal 2-day GOES-R L2 latency) to support critical space weather monitoring and prediction. 
\section{Open Research}
\subsection{Data Availability Statement}
SUVI L1b images used in this study can be found at \url{https://data.ngdc.noaa.gov/platforms/solar-space-observing-satellites/goes/goes16/l1b/}. The Levenberg-Marquardt algorithm used was applied with the python \href{https://lmfit.github.io/lmfit-py/}{LMFIT python package} \citep{newville_2015_11813}. NRLMSIS-2.0 outputs were produced using the \href{https://pypi.org/project/pymsis/}{pymsis NRLMSIS python-wrapper} \citep{Lucas2024-pz}. F$_{10.7}$ radio flux and Ap index data used as inputs to pymsis are obtained using the celestrak space weather python package, \url{https://pypi.org/project/spaceweather/}, with the data used collected by the GFZ German Research Centre for Geosciences \citep{Matzka2021-fk}. Provisional Dst data used in Figure \ref{fig:storm} was taken from the Kyoto Dst index service: \url{https://wdc.kugi.kyoto-u.ac.jp/dstdir/}.\add{ Color scheme for figures in this paper were adapted from \mbox{\cite{Tol2021}}}. 
SUVI solar occultation density and temperature profiles are available at \url{https://data.ngdc.noaa.gov/platforms/solar-space-observing-satellites/goes/goes16/l2/events/suvi_occultation/}.







\acknowledgments
This work was funded by NASA Space Weather Operations to Research (SWO2R) grant 80NSSC21K0028 and NASA's Heliophysics Flight Opportunities in Research and Technology (H-FORT) grant 80NSSC22K0343.


%
\bibliography{suvi_bib} 

\end{document}